\documentclass[12pt]{iopart}
\usepackage{graphicx}% Include figure files
\usepackage{amssymb}
\newcommand{\beq}{\begin{equation}}
\newcommand{\eeq}{\end{equation}}
\newcommand{\beqa}{\begin{eqnarray}}
\newcommand{\eeqa}{\end{eqnarray}}
\newcommand{\kB}{\mbox{$k_{\rm B}$}}
\newcommand{\kBT}{\mbox{$k_{\rm B}T$}}

\renewcommand{\vec}[1]{\mbox{\boldmath$#1$}}

\begin{document}

\title[Non-universal equilibrium crystal shape]{
%Quasi non-universal shape exponents induced by step-droplets around (001) facet on the equilibrium crystal shape:  The restricted solid-on-solid model with inter-step attraction of the point-contact type
%Non-universal shape exponents on the equilibrium crystal shape:  The restricted solid-on-solid model with inter-step attraction of the point-contact type
Non-universal equilibrium crystal shape results from sticky steps
}

\author{Noriko Akutsu}

\address{Faculty of Engineering, Osaka Electro-Communication
 University,\\
Hatsu-cho, Neyagawa, Osaka 572-8530, Japan}
\ead{nori@phys.osakac.ac.jp}
\begin{abstract}
The anisotropic surface free energy, Andreev surface free energy, and equilibrium crystal shape (ECS) $z=z(x,y)$ are calculated numerically using a transfer matrix approach with the density matrix renormalization group (DMRG) method.
The adopted surface model is a restricted solid-on-solid (RSOS) model with ``sticky'' steps, i.e., steps with a point-contact type attraction between them (p-RSOS model).
By analyzing the results, we obtain a first-order shape transition on the ECS profile around the (111) facet; and on the curved surface near the (001) facet edge, we obtain shape exponents having values different from those of the universal Gruber-Mullins-Pokrovsky-Talapov (GMPT) class.
In order to elucidate the origin of the non-universal shape exponents, we calculate the slope dependence of the mean step height of ``step droplets'' (bound states of steps) $\langle n (\vec{p})\rangle$ using the Monte Carlo method, where $\vec{p}=(\partial z/\partial x, \partial z/\partial y)$, and $\langle \cdot \rangle$ represents the thermal average.
Using the result of the $|\vec{p}|$ dependence of $\langle n (\vec{p})\rangle$, we derive a $|\vec{p}|$-expanded expression for the non-universal surface free energy $f_{\rm eff}(\vec{p})$, which contains quadratic terms with respect to $|\vec{p}|$.
The first-order shape transition and the non-universal shape exponents obtained by the DMRG calculations are reproduced thermodynamically from the non-universal surface free energy $f_{\rm eff}(\vec{p})$.

\end{abstract}

%Uncomment for PACS numbers title message
% Keywords required only for MST, PB, PMB, PM, JOA, JOB? 
%\vspace{2pc}
%\noindent{\it Keywords}: Article preparation, IOP journals
% Uncomment for Submitted to journal title message
%\submitto{\JPA}
% Comment out if separate title page not required
%
%added by N. Akutsu
\date{\today}
\maketitle

%\tableofcontents

\section{Introduction}

\begin{figure}%[!htb]
\begin{center}
\includegraphics[width=11 cm,clip]{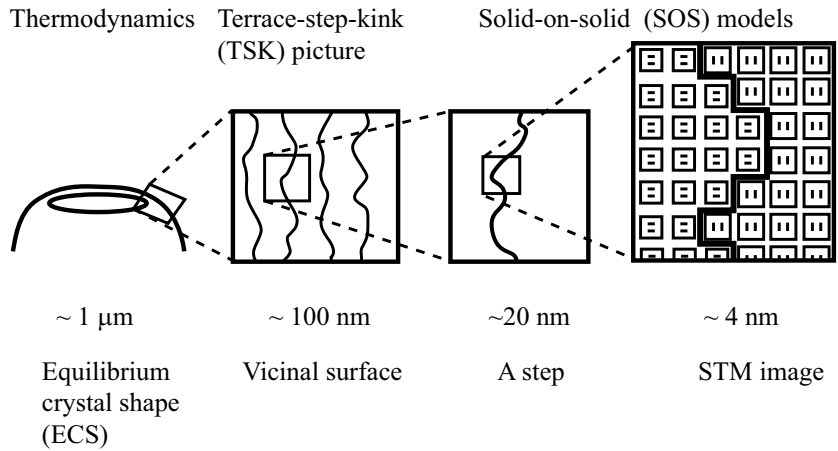}
\end{center}
\caption{\label{schematic}
Hierarchy of models depending on the length scale.
Example of a vicinal Si(001) surface. }
\end{figure}

The surface free energy is one of the most fundamental quantities in surface science.
The equilibrium crystal shape (ECS) (Fig. \ref{schematic}) is the shape of a crystal particulate with minimum surface free energy under equilibrium conditions\cite{wulff}-\cite{akutsu87-1}.
A polyhedral ECS is obtained using the Wulff theorem from a polar graph of the anisotropic surface free energy through the Wulff construction.
An ECS with several facets and curved areas is obtained using the Landau-Andreev method\cite{landau,andreev}, and this ECS is found to be similar to the Andreev surface free energy, where the work associated with step formation is eliminated from the surface free energy per projected area\cite{andreev}.
Since the ECS reflects the anisotropy of the surface free energy, studying the ECS corresponds to studying the surface free energy itself.

\begin{figure}%[!htb]
\begin{center}
\includegraphics[width=12 cm,clip]{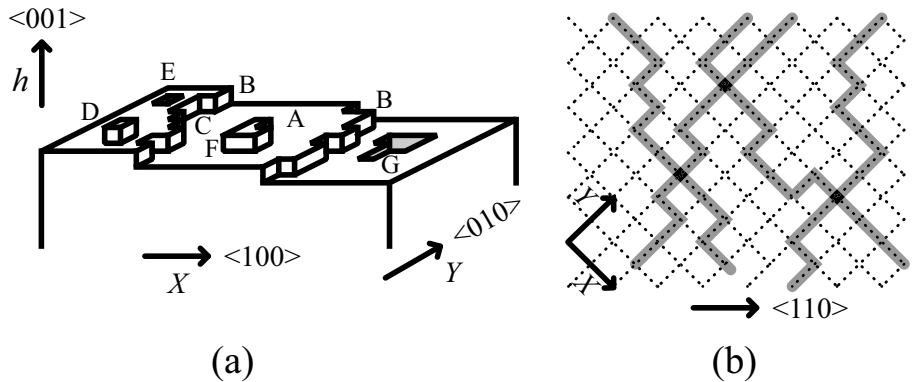}
\end{center}
\caption{\label{vicinal}
Vicinal surface represented by the restricted solid-on-solid (RSOS) model.
The difference in surface height between nearest neighbor sites is restricted to $\{0, \pm 1\}$. 
(a) Perspective view of surface tilted towards the $\langle 100 \rangle$ direction.
A: terrace.   B: step.   C: kinks. D: adatom.  E: ad-hole. 
F: island. G: negative island.
(b) Top view of surface tilted towards the $\langle 110 \rangle$ direction.
The gray lines represent surface steps and the filled squares represent points where adjacent steps collide.
}
\end{figure}

Recently, we applied the restricted solid-on-solid (RSOS) model coupled with the Ising system (RSOS-I model)\cite{akutsu99}-\cite{akutsu08} to investigate the interplay between surface steps and adsorbates on a vicinal surface.
The RSOS model\cite{rsos} (Fig. \ref{schematic} and Fig. \ref{vicinal}) is an SOS (or Kossel crystal)\cite{leamy,muller} model in which differences in height between nearest-neighbor (nn) surface sites are restricted to $\{0, \pm 1\}$. 
From statistical mechanical calculations using the RSOS-I model, a first-order shape transition is found to occur on the ECS profile.
In addition to the shape transition, the ``shape exponent'' on the ECS seems to have values different from the universal Gruber-Mullins-Pokrovsky-Talapov (GMPT) or one-dimensional (1D) free fermion values\cite{gmpt}-\cite{akutsu06}.  

Concerning the universal behavior of the GMPT, the universal form of the free energy is understood in terms of the terrace-step-kink (TSK) picture \cite{gmpt}-\cite{schultz} (Fig. \ref{schematic}) for a vicinal surface as follows:
\beq
f(\rho)= f(0)+\gamma \rho + B  \rho^3  + {\cal{O} }(\rho^4), \label{frho}
\eeq
where $\rho$ represents the step density, $\gamma$ represents the step tension, and $B$ represents the step interaction coefficient.
A many-body system of non-overlapping linear excitations embedded in two dimensions is known to have the form of free energy as expressed in Eq. (\ref{frho}).
The TSK picture  has been confirmed by exact calculations of the free energy using a microscopic body-centered cubic solid-on-solid (BCSOS) model\cite{beijeren,jayaprakash}.
Experimentally, the universal behavior expressed by Eq. (\ref{frho}) has been confirmed by observations of the ``shape exponent'' on the ECS profile \cite{ohachi}-\cite{nowicki}.
The ECS near a facet is expressed by $|z(x,y)-z(x_c,y_c)| = {\cal A} |x-x_c|^{3/2} $, where $(x_c,y_c)$ are the coordinates of the facet contour, the power on the right hand side is the shape exponent in the normal direction $\theta_n$, and ${\cal A}$ is the ``amplitude''. 
From Eq. (\ref{frho}), and following some thermodynamic calculations, the value of $\theta_n$ is obtained as $3/2$, which is a GMPT universal value.

In addition, based on recent developments in the study of non-equilibrium bunched steps\cite{stoyanov98-1}-\cite{misba10}, the values of the exponents in the profile of a bunched step are related to the force range of the effective step-step  interactions on the non-equilibrium vicinal surface.
The values of the exponents are thought to be related to the shape exponent $\theta_n$ on the ECS.
From these perspectives, the importance of studying the shape exponent is increasing.

In order to establish the non-GMPT shape exponent on the ECS, a $\rho$-expanded expression for surface free energy with non-GMPT terms should be derived.
In our previous work\cite{akutsu03}, we demonstrated the appearance of a short-range step-step attraction mediated by adsorbates, and we introduced the {\it step-droplet picture }\cite{akutsu03}.
Due to the complexity of interplay between surface steps and adsorbates, however, we could not derive a $\rho$-expanded expression for the non-GMPT surface free energy.

The aim of the present paper is to establish the non-GMPT shape exponent on the ECS. 
In other words, the purpose is to find a mechanism to obtain a $\rho$-expanded expression for surface free energy with non-GMPT terms in a system with a short-range step-step attraction.

To obtain clear results, we present a simple model: the RSOS model with a point-contact type step-step attraction (p-RSOS model)\cite{akutsu09,akutsu10}.
Physically, the step-step attraction represents the transient bond formed by the spatial overlap of orbitals between atoms at the collision point of the adjacent steps (Fig. \ref{vicinal}(b)).

The paper is organized as follows.
In \S 2, we present the definition of the p-RSOS model and show statistical mechanical calculations on the ECS, the equilibrium facet shape, and the non-universal shape exponents. 
In \S 3, a study on step droplets near equilibrium is described.
Calculation of the mean step height of the step droplets $\langle n \rangle$ is carried out using a Monte Carlo method.
In \S 4, we derive a $\rho$-expanded expression for the vicinal surface free energy, which contains non-GMPT terms.
In \S 5, using the non-GMPT vicinal surface free energy, we thermodynamically reproduce the results for the p-RSOS model obtained by the transfer matrix method in \S 2.
In \S 6, we present a summary and discussion.
Finally, a conclusion is given in \S 7.

\section{Statistical mechanical calculations using the p-RSOS model}

\subsection{Model Hamiltonian}

Let us consider the surface height $h(i,j)$ at a site $(i,j)$ on a square lattice to describe surface microscopic undulations (Fig. \ref{vicinal}).
In the RSOS model\cite{rsos}, the height differences between nearest-neighbor (nn) sites are restricted to values of $\{1,0,-1\}$.
We consider a point-contact type microscopic step-step interaction and refer to this model as the p-RSOS model.
The Hamiltonian for the p-RSOS model can then be written as
\beqa
{\cal H}_{\rm p-RSOS} &=& \sum_{i,j} \epsilon 
[ |h(i+1,j)-h(i,j)|+|h(i,j+1)-h(i,j)|]   \nonumber \\
&& +\sum_{i,j} \epsilon_{\rm int}[ \delta(|h(i+1,j+1)-h(i,j)|,2)  \nonumber \\
&& +\delta(|h(i+1,j-1)-h(i,j)|,2)] ,   \label{hamil}
\eeqa
where $\epsilon$ is the microscopic ledge energy, $\epsilon_{\rm int}$ is the microscopic step-step interaction energy, and $\delta(a,b)$ is Kronecker's delta.
The summation with respect to $(i,j)$ is performed all over sites on the square lattice.
The RSOS restriction is required implicitly.
In the case of $\epsilon_{\rm int}<0$, the interaction among steps becomes attractive.

For a vicinal surface, we add the terms of the Andreev field\cite{andreev} $\vec{\eta}= (\eta_x, \eta_y)$ to the Hamiltonian Eq. (\ref{hamil}) as an external field.
The model Hamiltonian given in Eq. (\ref{hamil}) for the vicinal surface then becomes
\beqa
{\cal H}_{\rm vicinal}&=&{\cal H}_{\rm p-RSOS} -\eta_x \sum_{i,j}[h(i+1,j)-h(i,j)] \nonumber \\
   &&-\eta_y \sum_{i,j}[h(i,j+1)-h(i,j)] .
\label{hamil_vicinal}
\eeqa 
The partition function ${\cal Z}$ for the p-RSOS model is given by
\beq
{\cal Z}= \sum_{\{h(i,j)\}} e^{-\beta {\cal H}_{\rm vicinal}} \label{partition}
\eeq
where $\beta= 1/\kBT$, $\kB$ is the Boltzmann constant, and $T$ is the temperature. 
The Andreev surface free energy $\tilde{f}(\vec{\eta})$ is the thermodynamic potential calculated from the partition function ${\cal Z}$ using
\beq
\beta \tilde{f}(\vec{\eta})= - \lim_{{\cal N} \rightarrow \infty} \frac{1}{{\cal N}} \  \ln {\cal Z}, \label{tildef}\\
\eeq
where ${\cal N}$ is the number of lattice points on the square lattice.
Practically, calculation of Eq. (\ref{tildef}) is not an easy task, because the entropy associated with the vast variety of zigzag structures of a surface step and by the parallel movement of steps is difficult to estimate.

Recently, the numerical renormalization group method for one-dimensional (1D) quantum spin systems has been further developed to become the density-matrix renormalization group (DMRG) method\cite{dmrg,dmrg2,natsume}.
Though the method is approximate and numerical, it successfully reproduces known exact  results with high precision.
One means of extending the DMRG method to classical systems is by mapping a two-dimensional (2D) classical system to a 1D quantum spin system\cite{dennij89} by use of the transfer matrix\cite{lieb72} together with the Suzuki-Trotter formula\cite{trotter}.
Such a method was developed by Nishino {\it et al.} for an infinite lattice, and is called the product-wave-function renormalization group (PWFRG) method \cite{pwfrg}-\cite{Ost-Rom}.
For the calculations in the present paper, we also adopt the PWFRG method.

\begin{figure}%[!htb]
\begin{center}
\includegraphics[width=11 cm,clip]{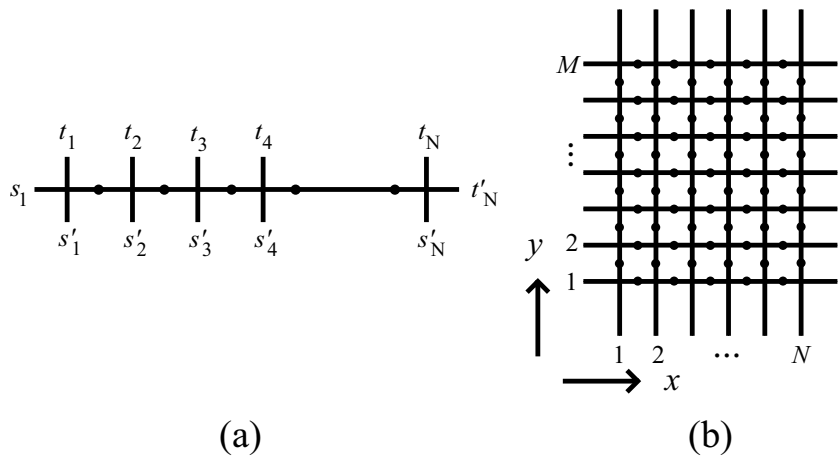}
\end{center}
\caption{\label{product}
(a) Schematic diagram of the transfer matrix.
(b) Graphical representation of the resultant matrix for the partition function.
}
\end{figure}

\begin{figure}%[!htb]
\begin{center}
\includegraphics[width=7.5 cm,clip]{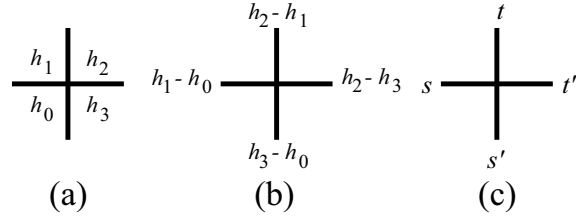}
\end{center}
\caption{\label{vertex1}
Mapping the solid-on-solid model onto the vertex model.
(a) Surface configuration of the SOS model.
(b) Mapping the configuration onto the vertex model.
(c) Vertex model.}
\end{figure}

\begin{figure}%[!htb]
\begin{center}
\includegraphics[width=10 cm,clip]{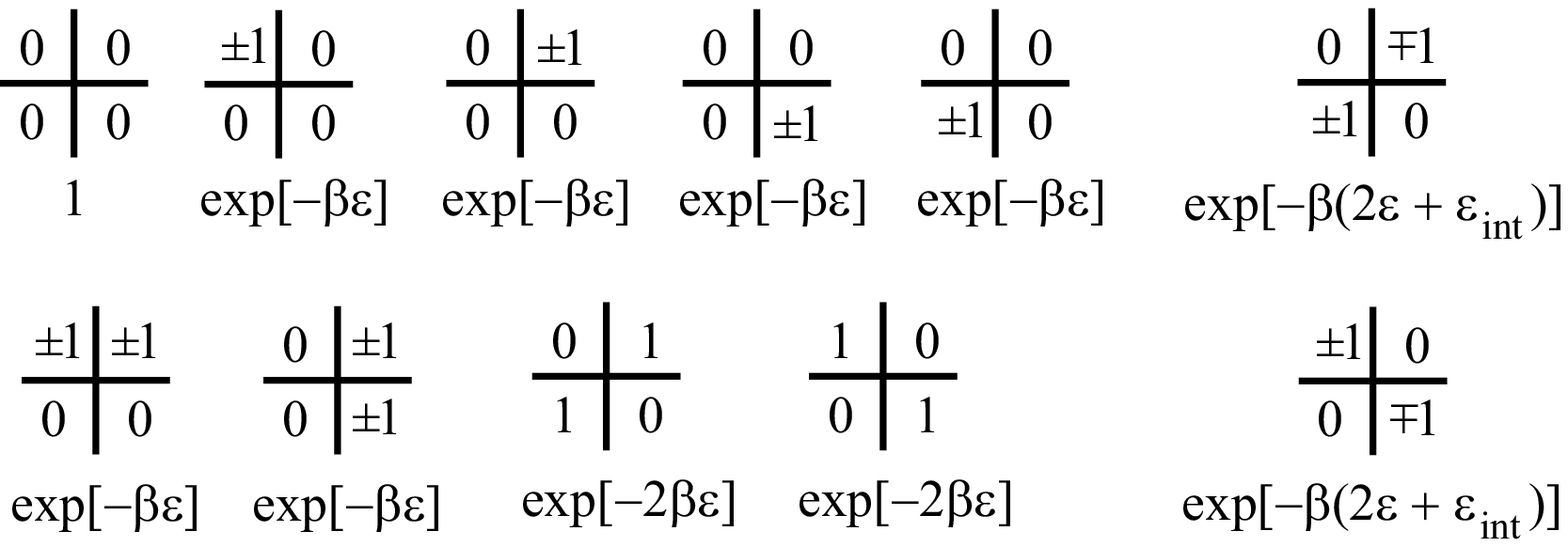}
\end{center}
\caption{\label{vertex3}
Explicit surface configurations corresponding to the 19 vertices and their statistical weights, where $\beta = 1/\kBT$.
}
\end{figure}

In order to apply the PWFRG method to the p-RSOS system, we construct the transfer matrix $\hat{T}(t_1,t_2,\cdots,t_N;s'_1,s'_2,\cdots,s'_N)$ (Fig. \ref{product}(a)) using a 19-vertex model\cite{akutsu98,dennij89} (Fig. \ref{vertex1}).
The partition function ${\cal Z}$ (Eq. (\ref{partition})) is then rewritten in terms of  $\hat{T}$ as
\beq
{\cal Z}= \Tr [\hat{T}(t_1,t_2,\cdots,t_N;s'_1,s'_2,\cdots,s'_N)^M] \label{partition2}
\eeq
where $N$ is the number of linked vertices and $M$ is the length of the system in the vertical direction in Fig. \ref{product}(b).
The statistical weight of each vertex is shown in Fig. \ref{vertex3}.
For a vicinal surface, the statistical weight is multiplied by $\exp[ \beta \{ (t+s')\eta_x + (s+t')\eta_y\}/2]$ with $s$, $t$, $s'$ and $t'$ having values of $\{0, \pm 1\}$.
Then, by use of the statistical weight denoted by $V(s,t;s',t')$, the transfer matrix is expressed as follows (Fig. \ref{product}(a)):  
\beqa
\hat{T}(t_1,t_2,\cdots,t_N;s'_1,s'_2,\cdots,s'_N) %\nonumber \\
&=& \sum_{\{q_1\},\{q_2\},\cdots}V(s_1,t_1;s'_1,q_1)V(q_1,t_2;s'_2,q_2)\nonumber \\
&&\cdots V(q_{N-1},t_N;s'_N,t'_N).
\eeqa
In the limit $M,N \rightarrow \infty$, only the largest eigenvalue of the transfer matrix $\Lambda(N)$ contributes to the partition function.
The Andreev surface free energy, therefore, is obtained from Eq. (\ref{tildef}) as
\beq
\beta \tilde{f}(\vec{\eta})= - \lim_{M,N \rightarrow \infty} \frac{1}{NM} \  \ln \Lambda(N) ^{M}. \label{tildeftm}
\eeq
The transfer matrix is diagonalized efficiently using the PWFRG method.
In the PWFRG calculation, the number of so-called ``retained bases'' $m$ is set from 7 to 37.
The number of iterations for the diagonalization process is set to $200 \sim 10^4$.

We also calculate the surface gradient $\vec{p}=(p_x,p_y)=(\partial z/\partial x, \partial z/\partial y)$ using the PWFRG method, where the surface gradient is defined as the thermal average of the height differences as follows:
\beqa
p_x(\vec{\eta}) &=&   \langle h(m+1,n)-h(m,n) \rangle d/a_x, \nonumber \\
p_y(\vec{\eta}) &=&   \langle h(m,n+1)-h(m,n) \rangle d/a_y. \label{pdef}
\eeqa
Here, $ \langle \cdot \rangle $ represents the thermal average, and $a_x$ and $a_y$ represent the lattice constants in the $x$ and $y$ directions, respectively ($d = a_x = a_y = 1$).
By sweeping the field $\beta \vec{\eta}$, we obtain curves for $p_x$ {\it vs.} $\beta \eta_x$ or  $p_y$ {\it vs.} $\beta \eta_y$\cite{akutsu99,akutsu03,akutsu06,akutsu09,akutsu10,akutsu98}.  

\subsection{First-order shape transition on the ECS profile\label{first_order}}
%\subsection{Andreev surface free energy}

\begin{figure}%[!htb]
\begin{center}
\includegraphics[width=17 cm,clip]{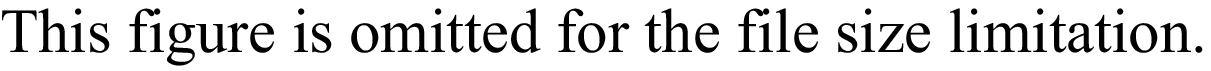}
\end{center}
\caption{\label{ECS3D}
Perspective views of the reduced ECS around the (001) facet calculated by the transfer matrix method with the PWFRG algorithm.
$(\beta \epsilon)^{-1}=\kBT / \epsilon=0.3$.
(a) p-RSOS model ($\epsilon_{\rm int}/\epsilon=-0.5$).
(b) The original RSOS model ($\epsilon_{\rm int}=0$).
}
\end{figure}

In Fig. \ref{ECS3D}, we show perspective views of the calculated Andreev surface free energy $\tilde{f}(\vec{\eta})$ divided by $\kBT$ around the (001) surface for $\kBT/\epsilon=0.3$. 
From the thermodynamics of the ECS, the Andreev surface free energy and the Andreev field are related to the ECS coordinates \cite{landau,andreev} by 
\beqa
\tilde{f}(\eta_x,\eta_y)= \lambda z(x,y), %\nonumber \\
\quad \eta_x = -\lambda x,
\quad \eta_y =-\lambda y, \nonumber \\
 p_x=-\frac{\partial \tilde{f}(\vec{\eta})}{\partial \eta_x}, 
\quad p_y=-\frac{\partial \tilde{f}(\vec{\eta})}{\partial \eta_y},  
\label{andreev_ecs}
\eeqa
where $\lambda $ is the Lagrange multiplier relating to the volume of the particulate.
Eq. (\ref{andreev_ecs}) implies that the surface shape of $\tilde{f}(\eta_x,\eta_y)$ is similar to the ECS.
We now introduce a reduced ECS $Z=Z(X,Y)$ such that $Z= \beta \tilde{f}(\eta_x,\eta_y)=\lambda \beta z(x,y)$, $X= \beta \eta_x= -\lambda \beta x $, and $Y= \beta \eta_y= -\lambda \beta y $.

In the case of $\epsilon_{\rm int} < 0$ (attractive step-step interaction), large $\{111\}$ facets appear in addition to the (001) facet and the $\{101\}$ facets, because the (111) surface is energetically stabilized by $\epsilon_{\rm int}$.
For comparison, we show the reduced ECS for the original RSOS model ($\epsilon_{\rm int}=0$) calculated by the PWFRG method in Fig. \ref{ECS3D}(b).

%\subsection{First-order transition on the ECS profile}

%\subsection{Reduced ECS for p-RSOS model}

\begin{figure}%[!htb]
\begin{center}
\includegraphics[width=9 cm,clip]{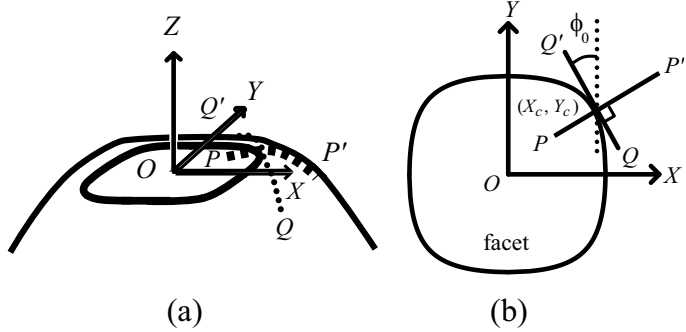}
\end{center}
\caption{\label{diagonal_line}
Normal line ($\overline{PP'}$) and tangential line ($\overline{QQ'}$) around the (001) facet on the reduced ECS. 
(a) Perspective view.
(b) Top view.
}
\end{figure}

\begin{figure}%[!htb]
\begin{center}
\includegraphics[width=13 cm,clip]{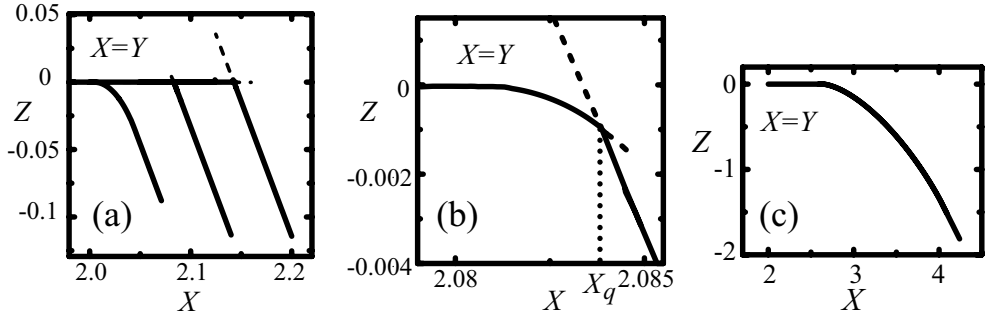}
\end{center}
\caption{\label{bfeta110}
Profile of the reduced ECS along the line $\overline{PP'}$.
$\phi_0=\pi/4$.
$\epsilon_{\rm int}/ \epsilon = -0.5$.
Broken lines represent metastable lines.
(a) From right to left, $\kBT/\epsilon= 0.35$, 0.36, and 0.37.
(b)  $\kBT/\epsilon= 0.36$.
The edge of the (111) facet is denoted by $X_q$.
(c) Original RSOS model ($\epsilon_{\rm int}=0$).
$\kBT/\epsilon= 0.3$.
}
\end{figure}

Along the line $\overline{PP'}$ shown in Fig. \ref{diagonal_line}, 
 we display the temperature dependence of the profile of the reduced ECS in Fig. \ref{bfeta110}.
As seen from the figure, the (001) facet ends at $(X_c,Y_c)$ and the curved region between the (001) and (111) facets represents a first-order shape transition at the edge of the (111) facet $(X_q,X_q)$\cite{akutsu09,akutsu10}.

\begin{figure}%[!htb]
\begin{center}
\includegraphics[width=10 cm,clip]{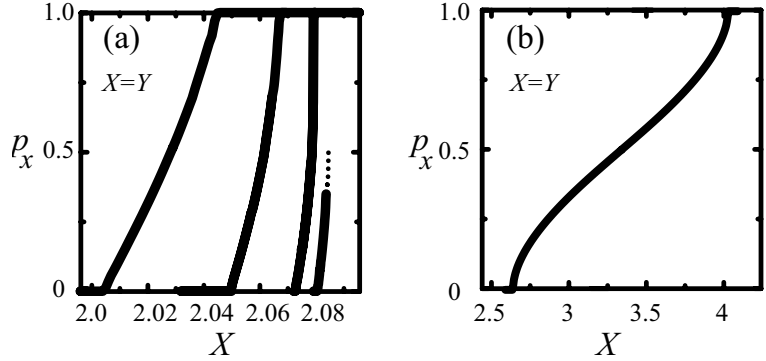}
\end{center}
\caption{\label{peta110}
$p_x$ {\it vs.} $X$ along the normal line $\overline{PP'}$ at  $\phi_0 = \pi/4$.
(a) $\epsilon_{\rm int}/ \epsilon = -0.5$.
From right to left, $\kBT/\epsilon= 0.36$, 0.361, 0.364, and 0.37.
The broken line indicates $p_x$ in the metastable state.
(b) Original RSOS model.  $\epsilon_{\rm int}=0$. $\kBT/\epsilon= 0.3$.
}
\end{figure}

Let us define the temperature $T_{f,1}$ as the highest temperature at which the first-order transition occurs at the (111) facet edge on the ECS profile.
We obtain $\kBT_{f,1}/\epsilon =0.3610 \pm 0.0005 $ for $\epsilon_{\rm int}/\epsilon = -0.5$ and $\phi_0=\pi/4$.
For temperatures $T \leq T_{f,1}$, the first-order transition occurs at $X_q(T)=Y_q(T)$ (Fig. \ref{bfeta110}(b)).
The surface slope changes abruptly from $p_0=1$ to $p_1$ at $X_q(T)$ for $0 \leq p_1<1$.
 In Fig. \ref{peta110}, we show $p_x(X,X)$ calculated by the PWFRG method along the line $\overline{PP'}$.
In the case of $\kBT/\epsilon= 0.36$, $p_x$ changes abruptly from $p_1=0.349 \pm 0.002$ to $p_0=1$ at $X_q = Y_q = 2.084$ ($> X_c=Y_c=2.0808 \pm 0.0002$).
The values of $p_x$ in the metastable state for $0.349< p_x < 0.501$ are shown by the broken line in the figure.

Let us define the temperature $T_{f,2}$ as the highest temperature at which the first-order transition occurs at the (001) facet edge. 
Below $T_{f,2}$, the curved area on the ECS between the (001) and (111) facets vanishes, and the (001) facet directly contacts the (111) facet.  
We obtain $\kBT_{f,2}/\epsilon =0.3585 \pm 0.0007$ for $\epsilon_{\rm int}/\epsilon = -0.5$ and $\phi_0=\pi/4$.

\subsection{Equilibrium facet shape\label{efs}}

\begin{figure}%[!htb]
\begin{center}
\includegraphics[width=12 cm,clip]{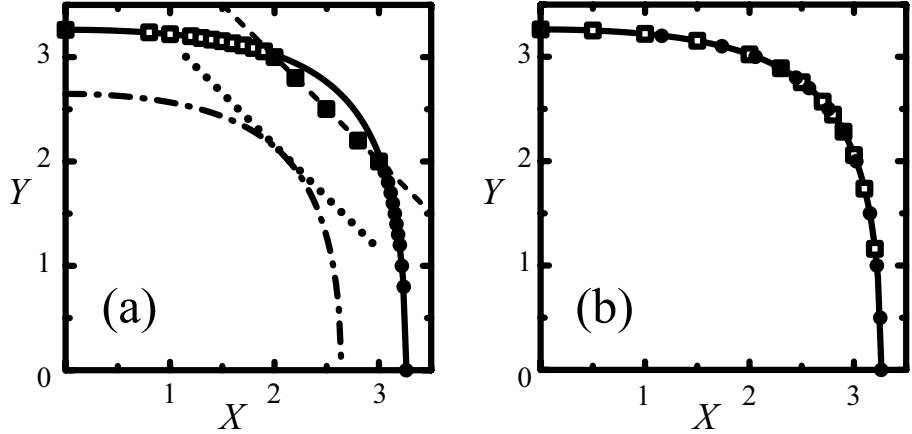}
\end{center}
\caption{\label{EFS}
Equilibrium facet shape (EFS) for $X>0$ and $Y>0$.
Filled circles: $(X_c,Y_c)$ values calculated by the PWFRG method for $\kBT/\epsilon = 0.3$.
Open squares: $(Y_c,X_c)$ values.
Solid lines: EFS for the 2D square nn Ising model (Eq. (\ref{ecs_Ising})) for $\kBT/\epsilon = 0.3$.
Dash-dotted lines:  EFS for the Ising model for $\kBT/\epsilon = 0.361$ (Eq. (\ref{ecs_Ising})).
Dashed lines: $Y=-X +5.0$.
Dotted lines: $Y=-X +4.1551$ (Eq. (\ref{intersection})).
(a) $\epsilon_{\rm int}/\epsilon=-0.5$.
(b) $\epsilon_{\rm int}=0$.
}
\end{figure}

In Fig. \ref{EFS}, we show the equilibrium facet shape (EFS) $(X_c,Y_c)$ of the (001) facet.
The filled circles represent the $(X_c,Y_c)$ values obtained from the contour of $Z(X,Y)=0$\cite{akutsu06} calculated by the PWFRG method.
Based on symmetry, we plot the $(Y_c,X_c)$ values as open squares.
The solid lines and dash-dotted lines in Fig. \ref{EFS} represent the exactly calculated 2D ECS for the interface of the 2D square nn Ising model\cite{akutsu86,rottman81,akutsu87-2} for $\kBT/\epsilon = 0.3$ and 0.361, respectively.
The exact expression for the 2D ECS for the nn square Ising model is given by the following equation\cite{akutsu90}-\cite{akutsu99iop} (\ref{2dising}):
\beq
\cosh (X_c)+\cosh (Y_c) = \frac{\cosh^2 (\beta \epsilon)}{\sinh (\beta \epsilon))}, \label{ecs_Ising}
\eeq
where $\beta = 1/\kBT$.
The dashed and the dotted lines in Fig. \ref{EFS}(a) represent the intersection between the (001) and (111) surfaces on the ECS for $\kBT/\epsilon = 0.3$ and 0.361, respectively. 
They are approximately calculated by the following equation: 
\beq
Y=-X + \beta (2\epsilon + \epsilon_{\rm int}). \label{intersection}
\eeq
%
%The PWFRG calculated values are coincide with the Eq. (\ref{EFS}) and Eq. (\ref{intersection}).
The facet shape obtained by the PWFRG calculations for $\kBT/\epsilon= 0.3$ agrees well with the 2D ECS for the 2D nn Ising model except for the part truncated by the intersection line between the (001) and (111) surfaces.

The step tension (or the interface tension) $\gamma (\phi) $ is given by\cite{akutsu06} (Eq. (\ref{tag8}))
\beq
\beta \gamma(\phi)= X_c \cos \phi +Y_c \sin \phi.  \label{eqgam}
\eeq
Therefore, the agreement between the EFS and the 2D ECS for the 2D Ising model also implies an agreement between the step quantities such as the step tension and the step stiffness ($\tilde{\gamma}(\phi)= \gamma(\phi) + \partial^2 \gamma(\phi)/\partial \phi^2$) for the RSOS model and the interface quantities such as the interface tension and the interface stiffness for the 2D Ising model\cite{akutsu86} (\ref{2dising}).
For $T<T_{f,2}$, the first-order shape transition occurs at the (001) facet edge.
In this case, a ``step'' actually corresponds to a ``giant step'' with a height $n d$ ($d=1$).
In the large step-height limit, $\lim_{n \rightarrow \infty} \gamma_{n}( \pi/4)/n$ converges to $\sqrt{2}(\epsilon + \epsilon_{\rm int}/2)$, which is smaller than $\gamma (\pi/4)_{\rm Ising}$.

Let us now calculate the approximate value of $T_{f,2}$.
For $T<T_{f,2}$, the EFS has the shape of the rounded square truncated by the intersection line  between the (001) and (111) surfaces, which is expressed by Eq. (\ref{intersection}).
For $T=T_{f,2}$, this line contacts the EFS at $(X_c^*, X_c^*)$ as the tangent line.
Then, from Eq. (\ref{ecs_Ising}) and (\ref{intersection}), we have 
\beq
\frac{\cosh^2 (\epsilon/\kBT_{f,2}^{(a)})}{\sinh (\epsilon/\kBT_{f,2}^{(a)})}  = 2 \cosh \left (\frac{2\epsilon+ \epsilon_{\rm int}}{2 \kBT_{f,2}^{(a)}}\right ).
\label{kBTf2}
\eeq
By solving Eq. (\ref{kBTf2}), we obtain $\epsilon/\kBT_{f,2}^{(a)} \approx 2.78778$ or $\kBT_{f,2}^{(a)}/\epsilon \approx 0.358709$, which is consistent with the PWFRG calculated value of $\kBT_{f,2}/\epsilon$ (\S \ref{first_order}).

\subsection{Non-universal shape exponents\label{non-univ-exponents}}

\begin{figure}%[!htb]
\begin{center}
\includegraphics[width=12 cm,clip]{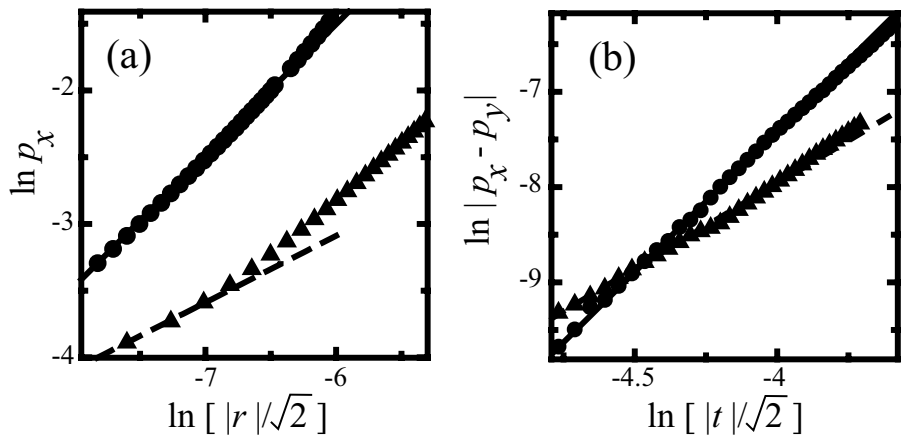}
\end{center}
\caption{\label{exponent}
Surface gradients calculated by the PWFRG method.
$\epsilon_{\rm int}/ \epsilon = -0.5$. 
$\phi_0 = \pi/4$
Filled circles: $\kBT/\epsilon= 0.36$.
Filled triangles: $\kBT/\epsilon= 0.37$.
$X_c=Y_c= 2.0808$ for $\kBT/\epsilon= 0.36$, and $X_c=Y_c= 2.0051$ for $\kBT/\epsilon= 0.37$, 
(a) $\ln(p_x)$ {\it vs.} $\ln [r/\sqrt{2}]$. 
Solid line: $\ln(p_x)=0.98 \ln  [r/\sqrt{2}] + 4.4$.
Broken line:  $\ln(p_x)=0.500 \ln  [r/\sqrt{2}]  -0.088$.
(b) $\ln |p_x-p_y| $ {\it vs.} $\ln [|t|/\sqrt{2}]$.
Solid line:  $\ln |p_x-p_y| =2.96 \ln [|t|/\sqrt{2}] + 4.4$
Broken line: $\ln |p_x-p_y| =1.8 \ln [|t|/\sqrt{2}] -0.79$.
}
\end{figure}

Let us assign $r$ and $\phi_0$ to the line $\overline{PP'}$ for $X>X_c$ so that $X= X_c+ r \cos \phi_0$ and $Y=Y_c+ r \sin \phi_0$, where $\phi_0=\pi/4$ is the tilt angle of the line $\overline{PP'}$ at $(X_c,Y_c)$ relative to the crystal axes (Fig. \ref{diagonal_line} (b)).
The ``normal shape exponent'' $\theta_n$ is defined as the shape exponent\cite{akutsu06} along $\overline{PP'}$ on the ECS profile such that $|Z(X(r),Y(r))-Z(X_c,Y_c)| = {\cal A}_n(\phi_0) r^{\theta_n}$ ($0 \leq r$), where we refer to the coefficient ${\cal A}_n(\phi_0)$ as the ``normal amplitude''.
Similarly, the ``tangential shape exponent'' $\theta_t$ and the ``tangential amplitude'' ${\cal A}_t(\phi_0)$  along $\overline{QQ'}$ are defined such that $|Z(X(t),Y(t))-Z(X_c,Y_c)| = {\cal A}_t(\phi_0) |t|^{\theta_t}$, where $t$ is a parameter assigned to the line $\overline{QQ'}$ as follows: $X= X_c-t \sin \phi_0,$ and $Y=Y_c+ t \cos \phi_0.$
Recalling that $\phi_0=\pi/4$, we express $(p_r, p_t)$ in terms of $(p_x,p_y)$ as follows:
\beq
p_r= \sqrt{2} p_x, \quad p_t= (p_y-p_x)/\sqrt{2}. \label{pr_pt}
\eeq
We show the logarithm of $p_r$ and $p_t$ in Fig. \ref{exponent}(a) and (b), respectively.

First, we study the shape exponents for $\kBT/\epsilon=0.36$, which is an example of $T_{f,2}< T <T_{f,1}$.
By fitting the data in the range $-7.8< \ln[r/\sqrt{2}]<-6.5$ in Fig. \ref{exponent}(a) to the linear function $A_0+A_1\ln[r/\sqrt{2}] $ by the least squares method, we obtain $A_1=0.98 \pm 0.03$, and $A_0=4.3 \pm 0.2$.
Similarly, by fitting the data in the range $-4.75< \ln[|t|/\sqrt{2}]<-4$ in Fig. \ref{exponent}(b) to $A_0'+A_1'\ln[|t|/\sqrt{2}] $, we obtain $A_1'=2.96 \pm 0.08$, and $A_0'=4.4 \pm 0.2$.
These values give the shape exponents and amplitudes as $\theta_n=1.98 \pm 0.03$, $\theta_t=3.96\pm 0.08$, ${\cal A}_n(\pi/4)= 40.6 \pm 0.2$, and ${\cal A}_t(\pi/4)= 7.2 \pm 1.0$.

Both exponents disagree with the GMPT universal values of $\theta_n=3/2$ and $\theta_t=3$\cite{akutsu06}.
Consequently, we conclude that the profile near the (001) facet contour for $T_{f,2}<T<T_{f,1}$ and $\phi_0=\pi/4$ shows non-GMPT behavior in the limit $p_r$, $p_t \rightarrow 0 $.

Next, we study the shape exponents for $\kBT/\epsilon=0.37$, which is an example of $T>T_{f,1}$.
As seen from Fig. \ref{exponent}(a), the slope of $\ln p_x$ crosses over from the larger value to the smaller value  as $\ln r$ decreases.
In Fig. \ref{exponent}(a), the broken line represents $\ln p_x= 0.5 \ln[r/\sqrt{2}] - 0.088$, and this line is determined by fitting to the three lowest points. 
Therefore, in the limit $r \rightarrow 0$, we have $\theta_n=1.5$ and ${\cal A}_n(\pi/4)= 0.73 \pm 0.04$.
For $\ln p_t$, we fitted the data in the range $-5< \ln[|t|/\sqrt{2}]< -4.3$ to $A_0'+ A_1' \ln[|t|/\sqrt{2}]$ and obtained $A_1' = 1.8 \pm 0.3$ and $A_0' = -0.79 \pm 0.08$.
These values lead to $\theta_t=2.8 \pm 0.3$ and ${\cal A}_t(\pi/4)= 0.062 \pm 0.018$.

From the values of $\theta_n$ and $\theta_t$ for $\kBT/\epsilon=0.37$, we conclude that the profile near the (001) facet contour for $T>T_{f,1}$ and $\phi_0=\pi/4$ behaves like the GMPT universal profile in the limit $p_r$, $p_t \rightarrow 0 $.

%\subsection{Non-universal amplitudes on the shape exponents\label{non-univ-amplitudes}}

\begin{figure}%[!htb]
\begin{center}
\includegraphics[width=14 cm,clip]{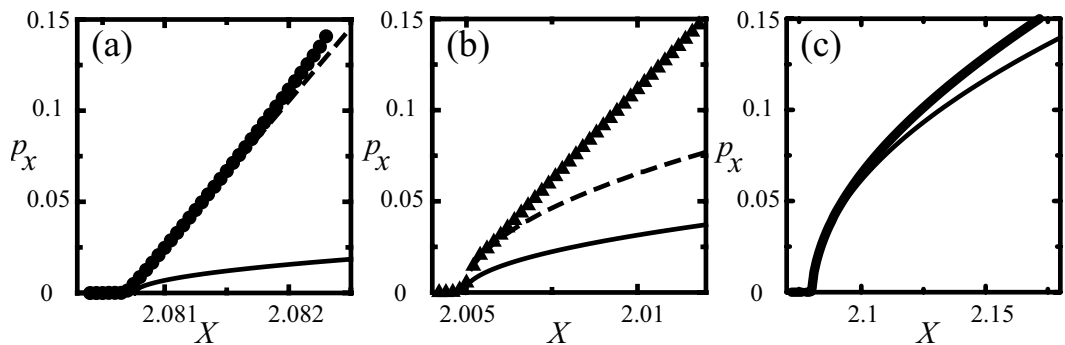}
\end{center}
\caption{\label{p_eta_detail}
$p_x$ {\it vs.} $X$ along the normal line $\overline{PP'}$.  $\phi_0 = \pi/4$.
(a) 
$\kBT/\epsilon= 0.36$.
$\epsilon_{\rm int}/ \epsilon = -0.5$.
$X_c=Y_c= 2.0808$.
Filled circles: PWFRG results.
Thin solid line: $p_x= 0.4432\sqrt{X-X_c}$ for $X>X_c$.
Broken line: $p_x= 80.4 X-167.3$
(b) 
$\kBT/\epsilon= 0.37$.
$\epsilon_{\rm int}/ \epsilon = -0.5$.
$X_c=Y_c= 2.0051$.
Filled triangles: PWFRG results.
Thin solid line: $p_x= 0.4421\sqrt{X-X_c}$ for $X>X_c$.
Broken line: $p_x= 0.9154\sqrt{X-X_c}$ for $X>X_c$. 
(c) 
Original RSOS model.  $\epsilon_{\rm int}=0$. 
$\kBT/\epsilon= 0.36$.
Thick solid line: PWFRG results.
Thin solid line: $p_x= 0.4432\sqrt{X-X_c}$ for $X>X_c$.
}
\end{figure}

%\subsection{Ising model calculation of the step tension and the step stiffness}

%GMPT amplitudes

We obtain the GMPT amplitudes using the equations\cite{akutsu06}
\beq
{\cal A}_n(\phi_0)= \frac{2\sqrt{2\beta \tilde{\gamma}(\phi_0)}}{3 \pi}, \quad
{\cal A}_t(\phi_0)= \frac{1}{3\pi \beta \tilde{\gamma}(\phi_0)},\label{amplitudes}
\eeq
where we use the universal relation\cite{akutsu88}
\beq
\beta  B(\phi_0)=\pi^2 /[6 \beta \tilde{\gamma}(\phi_0)].\label{univrel}
\eeq
In the temperature region $T \gtrsim T_{f,2}$, from the results shown in \S \ref{efs}, the step quantities of a single step is found to be well described by the interface quantities of the 2D nn Ising model. 
The exact expressions for the interface tension $\gamma(\phi)$ and the interface stiffness $\tilde{\gamma}(\phi)$ in the 2D nn Ising model for the case of $\phi_0=\pi/4$ are as follows\cite{akutsu86,rottman81,akutsu87-2}(Eq. (\ref{stiff_phi})):
\beqa 
\beta \gamma(\frac{\pi}{4})_{\rm Ising} &=& \sqrt{2}   \cosh^{-1} \left[ \frac{\cosh^2(\beta\epsilon)}{2\sinh(\beta\epsilon)} \right] ,  \nonumber \\
\beta \tilde{\gamma}(\frac{\pi}{4})_{\rm Ising} &=&  \sqrt{2}  \tanh \left[\beta  \gamma(\frac{\pi}{4})_{\rm Ising}/\sqrt{2} \right]  \label{stiff11}. 
\eeqa
For $\kBT/\epsilon=(\beta \epsilon)^{-1}= 0.36$, we have $\beta \gamma(\pi/4)_{\rm Ising}/\sqrt{2} \approx 2.08075$, and $\beta \tilde{\gamma}(\pi/4)_{\rm Ising} \approx  1.371$, and at $\kBT/\epsilon=0.37$, we have $\beta \gamma(\pi/4)_{\rm Ising}/\sqrt{2}  \approx  2.0051$, and $\beta \tilde{\gamma}(\pi/4)_{\rm Ising}\approx  1.364$.
Using these values for the interface stiffness, we can obtain the GMPT amplitudes from Eq.(\ref{amplitudes}).

In Fig. \ref{p_eta_detail}, we show $p_x(r)$  along the normal line $\overline{PP'}$.
The solid curve in Fig. \ref{p_eta_detail} shows $p_x(r)$ calculated from Eq. (\ref{pr_pt}) together with the GMPT amplitude determined using Eq. (\ref{amplitudes}).
As seen from the figure, the solid curve disagrees with the values calculated by the PWFRG method not only for $\kBT/\epsilon = 0.36$ (Fig. \ref{p_eta_detail}(a)) but also for $\kBT/\epsilon = 0.37$ (Fig. \ref{p_eta_detail}(b)).
For $\kBT/\epsilon=0.36$, with $p_x<0.08$, fitting the values obtained by the PWFRG method to $p_x = A_0 + A_1 X$ gives $A_0=167.3 \pm 0.3$ and $A_1=80.4 \pm 0.2$.
This is shown as the broken line in Fig. \ref{p_eta_detail}(a).
For $\kBT/\epsilon=0.37$, the broken line in Fig. \ref{p_eta_detail}(b) represents $p_x= 0.9154\sqrt{X-X_c}$, based on the value of $A_0'$ obtained from the straight line in Fig. \ref{exponent}(a). 
The curve agrees with $p_x$ calculated by the PWFRG method for small $X-X_c$.

\subsection{Vicinal surface tilted towards the $\langle 100 \rangle$ direction}

\begin{figure}%[!htb]
\begin{center}
\includegraphics[width=13 cm,clip]{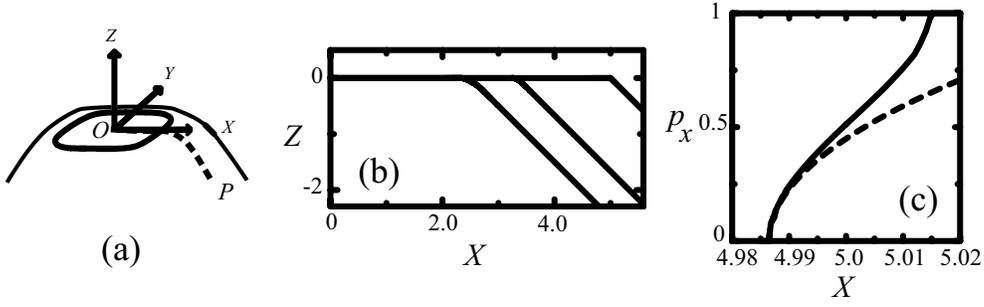}
\end{center}
\caption{\label{bfeta100}
Profile of the reduced ECS along the curve $\overline{OP}$ in the p-RSOS model.
$\epsilon_{\rm int}/ \epsilon = -0.5$.
(a) Schematic diagram of the ECS and the curve $\overline{OP}$ (broken line).
(b) $\tilde{f}/\kBT$ ($Z$) {\it vs.} $X$. 
From right to left, $\kBT/\epsilon= 0.2$, 0.3, and 0.4.
(c) Surface slope $p_x$ {\it vs.} $X$.
$\kBT/\epsilon = 0.2$.
$X_c=4.9865$, $Y_c=0$.
Broken line: $p_x$ calculated using Eq. (\ref{phi0}) with Eq. (\ref{stiff10}).
}
\end{figure}

Let us now consider the profile along the curve $\overline{OP}$ in Fig. \ref{bfeta100}(a).
We show the calculated Andreev surface free energy and the $p_x$-$X$ curve in Fig. \ref{bfeta100}(b) and (c), respectively.
For $\phi_0=0$, we obtain $p_x$ and $p_y$ from Eq. (\ref{andreev_ecs}) and  Eq. (\ref{amplitudes}) and the GMPT universal shape exponents as
\beq
p_x= \frac{\sqrt{2\beta \tilde{\gamma }(0)}}{\pi }(X-X_c)^{1/2}|_{Y=0}, \quad 
p_y = \frac{1}{\pi \beta \tilde{\gamma}(0)}Y^2|_{X=X_c},\label{phi0}
\eeq
where $Y_c=0$ and $X_c=4.9865 \pm 0.0005$.
The interface tension and stiffness in the Ising model are obtained exactly from Eq. (\ref{eqgam}) and Eq. (\ref{stiff_phi}) as follows:
\beqa 
\beta \gamma(0)_{\rm Ising} =   \cosh^{-1} \left[ \frac{\cosh^2(\beta \epsilon )}{\sinh(\beta \epsilon )} -1  \right] , \nonumber \\
\beta \tilde{\gamma}(0)_{\rm Ising} =    \sinh \left[ \beta \gamma(0)_{\rm Ising}  \right]  \label{stiff10}. 
\eeqa
The value of $X_c$ agrees well with $\beta \gamma(0)_{\rm Ising}$.
$p_x$ calculated using Eq. (\ref{phi0}) with Eq. (\ref{stiff10}) is plotted as the broken line in Fig. \ref{bfeta100}(c), and it can be seen that it closely matches the $p_x$-$X$ curve calculated by the PWFRG method for small $p_x$.

Therefore, we conclude that the vicinal surface tilted towards the $\langle 100 \rangle$ direction shows typical GMPT universal behavior.  

\section{Origin of non-universal behavior: Step droplets \label{origin}}

\subsection{Vicinal surface free energy}

When the surface slope is chosen to be an external variable instead of the Andreev field $\vec{\eta}$, the thermodynamic function of the surface becomes the vicinal surface free energy $f(\vec{p})$.  
Using the relationship between the Andreev surface free energy and the vicinal surface free energy\cite{andreev}, we obtain $f(\vec{p})$ from $\tilde{f}(\vec{\eta})$ as
\beq
f(\vec{p})= \tilde{f}(\vec{\eta})+ \vec{\eta} \cdot \vec{p}. 
 \label{legendre}
\eeq
Namely, using the notations on the reduced ECS, we have
\beq
\beta f(\vec{p}) = Z+ Xp_x+Yp_y .
 \label{legendre2}
\eeq

\begin{figure}%[!htb]
\begin{center}
\includegraphics[width=9 cm,clip]{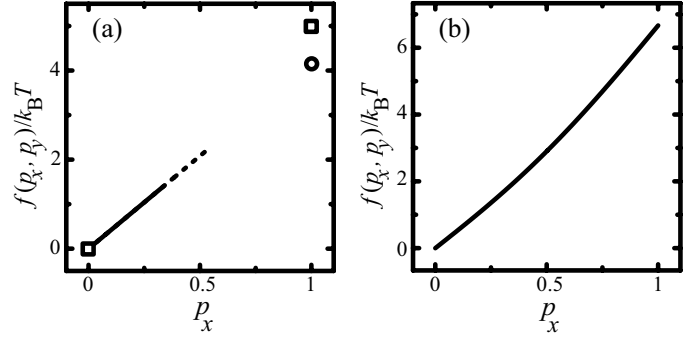}
\end{center}
\caption{\label{bfpp}
Surface free energy per projected area $f(\vec{p})$ (vicinal surface free energy).
(a) p-RSOS model.  $\epsilon_{\rm int}/\epsilon=-0.5$.  $p_x=p_y=p$.
Solid line: $f(p,p)$ at $\kBT/\epsilon = 0.36$.
Dashed line: $f(p,p)$ at $\kBT/\epsilon = 0.36$ for the metastable state.
Open circle: $f(1,1)$ at $\kBT/\epsilon = 0.36$.
Open squares: $f(0,0)$ and $f(1,1)$ at $\kBT/\epsilon = 0.3$.
(b) The original RSOS model ($\epsilon_{\rm int}=0$). 
Solid line: $f(p,p)$ at $\kBT/\epsilon = 0.3$.
}
\end{figure}

In Fig. \ref{bfpp}(a), $f(\vec{p})$ in the p-RSOS model ($\epsilon_{\rm int}/\epsilon = -0.5$) calculated by Eq. (\ref{legendre2}) is shown, and $f(\vec{p})$ in the original RSOS model is shown in Fig. \ref{bfpp}(b).
For $\kBT/\epsilon = 0.3$, only $f(0,0)$ and $f(1,1)$ are obtained by the PWFRG calculations,  because a vicinal surface with a regular train of steps in the region $0<p<1$ is thermodynamically unstable. 
For $\kBT/\epsilon = 0.36$, a $f(\vec{p})$ curve is obtained in the region $0\leq p_x \leq 0.349$, and a curve for the metastable state in the region $0.349 <p_x <0.501$.
For $\kBT/\epsilon = 0.3$ and 0.36, $f(1,1)$ is well approximated by $f(1,1)=2 \epsilon + \epsilon_{\rm int}$.

\subsection{Thermal step bunching\label{bunching}}

\begin{figure}%[!htb]
\begin{center}
\includegraphics[width=12 cm,clip]{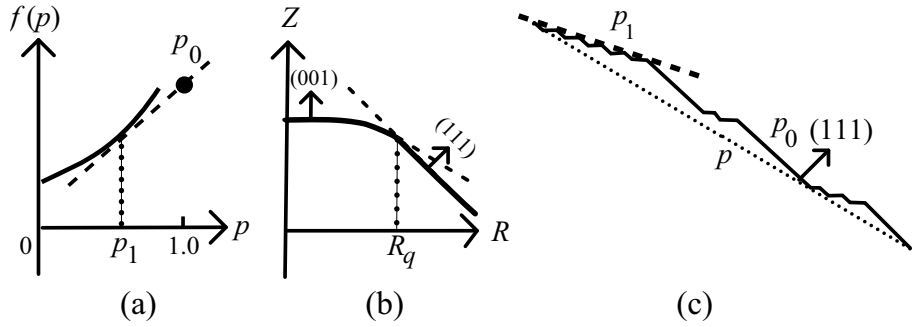}
\end{center}
\caption{\label{fig_hill_and_valley08}
Thermal step bunching and coexistence of two surfaces.  
(a) Schematic diagram of vicinal surface free energy $f(p)\equiv f(p,p)$.
Filled circle: $f(p_0)=f(1)$.
Broken line: tangential line connecting $(p_1,f(p_1))$ and $(p_0,f(p_0))$ with a slope of $R_q$ (Eq. (\ref{co-tangent}))\cite{akutsu03}.
(b) Schematic diagram of profile of the reduced ECS. 
$R=\sqrt{X^2+Y^2}$.
Broken line: shape in the metastable state.
The (111) facet edge, where the first-order transition occurs, is indicated by $R_q$.
A surface with a slope $p_0$ coexists with a surface with a slope $p_1$ at $R_q$.
(c) Step bunching when two surfaces coexist. 
Dotted line: mean surface slope ($= p$).
Broken line: local surface with slope $p_1$.
}
\end{figure}

The first-order shape transition on the ECS profile around the (111) facet (Fig. \ref{fig_hill_and_valley08}(b)) leads to the coexistence of two surfaces in equilibrium (Fig. \ref{fig_hill_and_valley08}(a))\cite{akutsu03,akutsu08,akutsu09}, and we refer to this process as thermal step bunching\cite{akutsu99,akutsu03}.
Let us consider the free energy $\bar{f}(p)$ ($p=|\vec{p}|$) along $\overline{PP'}$ in Fig. \ref{diagonal_line} for a mixture of surfaces with slopes of $p_0$ and $p_1$ so that the mean slope is $p$, as shown in Fig. \ref{fig_hill_and_valley08}(c).
The free energy of the simple mixed system is described as 
\beqa
\bar{f}(p)=\bar{x}_1 f(p_1)+ \bar{x}_0 f(p_0) \nonumber \label{fpp1}\\
\bar{x}_1 + \bar{x}_0 =1
\eeqa
where $\bar{x}_0$ and $\bar{x}_1$ represent the fractional areas of the surfaces with slopes of $p_0$ and $p_1$, respectively. Since $\bar{x}_0=(p-p_1)/(p_0-p_1)$ and $\bar{x}_1=(p_0-p)/(p_0-p_1)$,
$\bar{f}(p)$ can be rewritten as
\beq
\bar{f}(p)=f(p_1)+\frac{[f(p_0) - f(p_1)]}{p_0-p_1}(p-p_1). \label{co-tangent}
\eeq
Eq. (\ref{co-tangent}) describes the co-tangent line which contacts $f(p)$ at points $(p_0, f(p_0))$ and $(p_1, f(p_1))$ (Fig. \ref{fig_hill_and_valley08}(a)).
For $p_1<p<p_0$, therefore, the free energy of the mixed surfaces is lower than that for a homogeneous surface.
Moreover, since $R= \beta \partial f(\vec{p})/\partial p_r$,  Eq. (\ref{co-tangent}) is equivalent to $Z(R_q)=\beta \tilde{\bar{f}}(R_q)=\beta \bar{f}(p)- R_q p = \beta \tilde{f}(R)|_{R \rightarrow R_{q},-} = \beta \tilde{f}(R)|_{R \rightarrow R_{q},+} $, where $\beta= 1/\kBT$ (Fig. \ref{fig_hill_and_valley08}(b)).

\subsection{Mean step height\label{mean}}

\subsubsection{Giant steps: $T< T_{f,2}$\label{faceting}}

\begin{figure}%[!htb]
\begin{center}
\includegraphics[width=11 cm,clip]{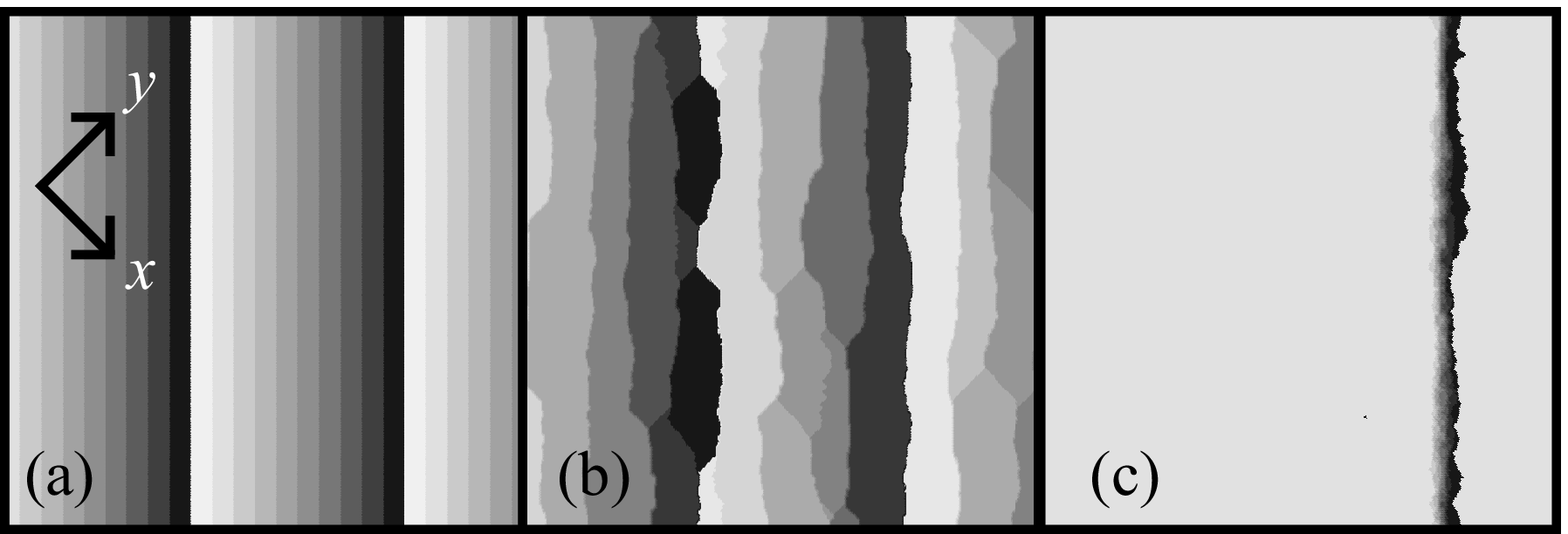}
\end{center}
\caption{\label{mcsfigLowT}
Snapshots of the simulated vicinal surfaces.
Top view.
$\epsilon_{\rm int}/\epsilon=-0.5$.

$240 \sqrt{2} \times 240 \sqrt{2}$.
Surface height is represented by brightness with 10 gradations, with brighter regions being higher.
The black areas next to the white areas represent the higher terraces than the ones by one because of the finite gradation.  
(a) Initial configuration. Number of steps = 24.
(b) $\kBT/\epsilon = 0.1$.   $5 \times 10^6$ MCS/site. (c) $\kBT/\epsilon = 0.35$ (number of steps = 10).   $4 \times 10^8$ MCS/site. 
}
\end{figure}

We consider that the origin of the non-GMPT shape exponents is the formation of local ``step droplets''\cite{akutsu99,akutsu03}, similar to the formation of clusters in a gaseous system near the transition temperature.
In order to form a clear image of such step droplets, we investigate the step dynamics near equilibrium using a Monte Carlo method for a vicinal surface tilted towards the $\langle 110 \rangle $ direction at low temperature.

Initially, steps numbering $N_{\rm step}$ are spaced evenly on a surface with an area of $240 \sqrt{2} \times 240 \sqrt{2} $ (Fig. \ref{mcsfigLowT}(a)). 
The left side of the image is higher than the right side by an amount equal to $N_{\rm step}$.
Periodic boundary conditions are imposed in the vertical direction in Fig. \ref{mcsfigLowT}.

To study the time evolution of the step configuration, we adopt a simple Metropolis algorithm without any driving force to simulate crystal growth.
We randomly choose a site $(i,j)$, and allow its height $h(i,j)$ to increase or decrease with equal probability. Then, if the RSOS restriction is satisfied, the height is updated by the Metropolis algorithm with a probability $P$ described by
\beq
P=\left\{
\begin{array}{ll}
1&   ( \Delta E(i,j)\leq 0 ),  \\
\exp[- \beta \Delta E(i,j)] & ( \Delta E(i,j)>0 ),
\end{array}
\right.
\label{prob}
\eeq
where $\Delta E(i,j)= E(h(i,j)\pm 1)- E(h(i,j))$.
The energy $E(h(i,j))$ is calculated using the p-RSOS Hamiltonian shown in Eq. (\ref{hamil}).

In the present Monte Carlo simulation, only non-conserved attachments and detachments of atoms are taken into consideration.
Other effects that occur on a real surface are ignored, such as surface diffusion\cite{bcf}, electromigration\cite{stoyanov}-\cite{uwaha}, the Schwoebel effect\cite{pimpinelli,weeks}, impurity effects\cite{weeks94}-\cite{frank}, strain effects\cite{teichert}-\cite{ibach}, and the effect of surfactants\cite{vonhoegen98}-\cite{minoda}.

For $T < T_{f,2}$, Fig. \ref{mcsfigLowT} shows the formation of giant steps, similar to ``step faceting''\cite{mullins,herring}.
As mentioned in our previous paper\cite{akutsu10}, at sufficiently low temperatures, step zipping occurs (Fig. \ref{mcsfigLowT}(b)), whereas step unzipping seldom takes place.
The zipping process corresponds to successive sticking together of steps, starting from the colliding point of adjacent steps, similar to the action of a zip fastener.
At slightly higher temperatures where unzipping occurs more frequently, the vicinal surface reaches an equilibrium configuration (Fig. \ref{mcsfigLowT}(c)).
All the steps join together to form a single giant step, whose edge has the appearance of a smooth (111) surface. 
\subsubsection{Step droplets: $T_{f,2} < T \leq T_{f,1}$}

\begin{figure}%[!htb]
\begin{center}
\includegraphics[width=6 cm,clip]{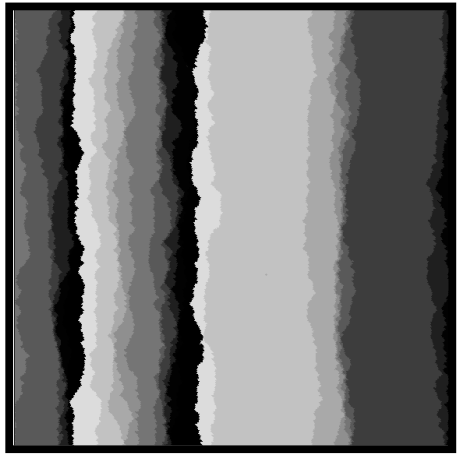}
\end{center}
\caption{\label{mcsfig}
Snapshot of simulated vicinal surfaces.
Top view.
$\epsilon_{\rm int}/\epsilon=-0.5$.
$4 \times 10^8$ MCS/cite.
$240 \sqrt{2} \times 240 \sqrt{2}$.
Number of steps: 24.
Surface height is represented by brightness with 10 gradations, with brighter regions being higher.
$\kBT/\epsilon = 0.36$.
}
\end{figure}

In this temperature regime, the fluctuations on the step meandering become larger. 
A snapshot of the surface for $\kBT/\epsilon=0.36$ is shown in Fig. \ref{mcsfig}. 
In contrast to the low-temperature surface structure, giant steps are seen to exist locally, and become larger and smaller dynamically.
We refer to such local giant steps as ``step droplets''. Based on the structure of giant steps at low temperature, we now use Monte Carlo simulations to determine the size of step droplets and the mean number of steps $\langle n \rangle$ contained in giant steps. 
\begin{figure}%[!htb]
\begin{center}
\includegraphics[width=9 cm,clip]{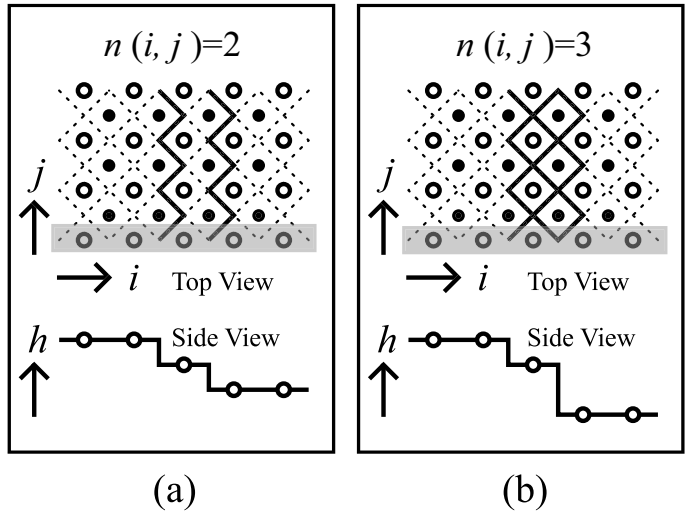}
\end{center}
\caption{\label{nstep}
Schematic illustration of bunched step.
Open circles: A-sublattice points.
Closed circles: B-sublattice points.
Upper figure: top view.
Bottom figure: side view of the $j$-th row (shaded area) in the top view.
Solid line (top view): surface steps.
(a) $n(i,j)=2$.
(b) $n(i,j)=3$.
}
\end{figure}

In order to calculate $\langle n \rangle$, we divide the square lattice into two sub-lattices, which are represented by open circles and closed circles in Fig. \ref{nstep}, respectively.
Sweeping $i$ from left to right for fixed $j$ on the A-sublattice, we define the beginning of a step droplet at $(i,j)$ as the position where successive height changes begin to occur.
If these successive height changes ends at $(i+n,j)$ in the A-sublattice, the size of the step droplet is taken to be $n(i,j)$.
The mean size of the step droplets $\langle n \rangle$ is then defined by
\beq
\langle n \rangle =\frac{1}{N_{\rm droplet}} \sum_{j=1}^{N_j}\sum_{i=1}^{N_i}|n(i,j)|, \label{def_n_av}
\eeq
where $N_i=N_j=240$ is the number of lattice units in the $i$- and $j$- directions, respectively.
$N_{\rm droplet}$ is calculated using $N_{\rm droplet}=\sum_{j=1}^{N_j} N_{\rm droplet}(j)$, where $N_{\rm droplet}(j)$ represents the number of giant steps (step droplets) in the $j$-th row of the square lattice shown in Fig. \ref{nstep}.

\begin{figure}%[!htb]
\begin{center}
\includegraphics[width=12 cm,clip]{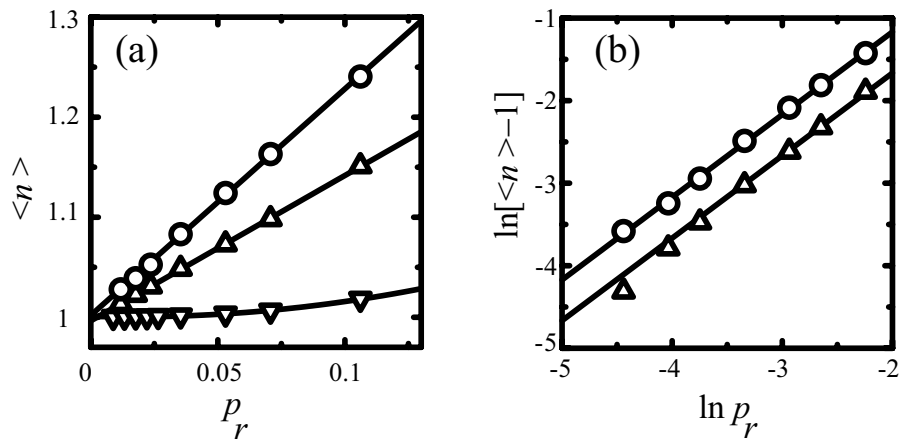}
\end{center}
\caption{\label{nfig}
Slope dependence of mean step-droplet size.
Monte Carlo calculation.
Lattice size: $240 \sqrt{2} \times 240 \sqrt{2}$.
$\epsilon_{\rm int}/\epsilon= -0.5$.
 Open circles: $\kBT/\epsilon = 0.36$, $8 \times 10^8$ MCS/site. 
Open triangles: $\kBT/\epsilon = 0.37$, $4 \times 10^8$ MCS/site.
Open reversed triangles: $\kBT/\epsilon = 0.36$, $\epsilon_{\rm int}=0$ (original RSOS model), $2 \times 10^8$ MCS/site.
}
\end{figure}

In Fig. \ref{nfig}(a), we plot $\langle n \rangle$ calculated using the Monte Carlo method against $p_r$, as expressed in Eq. (\ref{pr_pt}).
As can be seen, $\langle n \rangle$ increases linearly with $p_r$.
Fitting the data to $\langle n \rangle =A_0 + A_1 p_r$ using the least squares method, we obtain the values $A_0= 1.00 \pm 0.02$ and $A_1= 2.26 \pm 0.08$.
To confirm the linearity, $\ln[\langle n \rangle -1]$ {\it vs.} $\ln p_r$ is plotted in Fig. \ref{nfig}(b).
Fitting the data for $p_r >0.03$ to $ \ln [\langle n \rangle -1] = A_0+A_1 \ln p_r $ yields $A_0= 0.84 \pm 0.04$ and $A_1= 1.00 \pm 0.02$.
The upper fitted line in Fig. \ref{nfig}(b) is seen to agree with the data for $p_r <0.03$ (open circles).

\subsubsection{$T > T_{f,1}$}

In this temperature regime, the GMPT universal behavior is expected.
As seen from Fig. \ref{nfig}, however, the GMPT universal behavior such as that for the original RSOS model (reversed open triangles) is not observed for $\kBT/\epsilon = 0.37$ (open triangles).

Fitting the data for the original RSOS model to $\langle n \rangle = A_0+A_1 p_r +A_2 p_r^2$ yields $A_0 = 1.001 \pm 0.008$, $A_1= -0.07 \pm 0.09$, and $A_2= 2.2 \pm 1.0$.
These values suggest that $A_0=1$ and $A_1=0$ for the GMPT universal system in the limit $p_r \rightarrow 0$. 

For the p-RSOS model, $\langle n \rangle$ seems to increase linearly with $p_r$ for $\kBT/\epsilon =0.37$.
Fitting the data for $\kBT/\epsilon = 0.37$ in Fig. \ref{nfig}(a) to $\langle n \rangle = A_0' + A_1' p_r$ yields $A_0'=0.996 \pm 0.04$ and $A_1' = 1.45 \pm 0.03$.
Fitting the data for $\kBT/\epsilon = 0.37$ and $p_r> 0.03$ in Fig. \ref{nfig}(b) to $\ln (\langle n \rangle-1)=A_0'' + A_1'' \ln p_r$ yields $A_0''=0.58 \pm 0.5$ and $A_1'' = 1.09 \pm 0.07$.

\section{$\vec{p}$-expanded expression for non-GMPT vicinal surface free energy}

\subsection{Breakdown of homogeneous 1D fermion system}

Let us consider the mean running direction of steps as the direction of time flow.
If we consider a coarse-grained vicinal surface of the original RSOS model, the steps can be regarded as continuous lines describing the space-time trajectories of Brownian particles with hard-core repulsion.
If the particle density is low, the Hamiltonian for the transfer matrix in the continuous system is known to be described by 1D free fermions\cite{villain}-\cite{schultz},\cite{akutsu88}, and the free energy is expressed by Eq. (\ref{frho}).

In the case of $\epsilon_{\rm int} \neq 0$, since two fermions can not be present at the same site at the same time, the transfer matrix for the p-RSOS model for a continuous system cannot be written as the Hamiltonian for interacting fermions.
Using the transfer matrix of the RSOS model, which is similar to the p-RSOS model, den Nijs and Rommels mapped the RSOS model to a 1D quantum spin system\cite{dennij89}. 
Since the domain walls in the ordered phase correspond to the steps on the surface, the Hamiltonian for the transfer matrix in the continuous system is expressed in terms of interacting impenetrable bosons\cite{holstein,yamamoto10}.

In the case of $\epsilon_{\rm int}<0$, bound states such as molecules or clusters might be formed between nn bosons\cite{messia} depending on the strength of the attraction between the bosons.
That is, $\gamma$ and $B$ in Eq. (\ref{frho}) can vary depending on the properties of the bound state,  such as the number of particles in a cluster.
In contrast, in a 1D free fermion system, since $\gamma$ and $B$ are the microscopic quantities assigned to a fermion monomer in the continuous model, changes in these parameters are not allowed.
In this way, the 1D fermion picture breaks down, and we must take inhomogeneous effects such as cluster formation in the Bose gas into consideration (a similar argument is given in \S \ref{bunching}).

\subsection{Expression for non-GMPT vicinal surface free energy\label{non-gmpt}}

\begin{figure}%[!htb]
\begin{center}
\includegraphics[width=9 cm,clip]{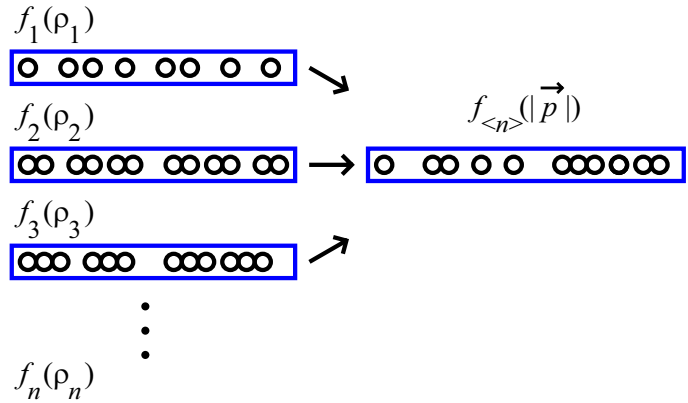}
\end{center}
\caption{\label{fermions}
Schematic illustration of 1D many-body systems with boson $n$-mers, and their mixture. (see \S \ref{non-gmpt}.)
}
\end{figure}

We regard a giant step with a size $n$ as a boson $n$-mer formed by microscopic step-step attraction $\epsilon_{\rm int}$ ($<0$).
As time passes, these $n$-mers may grow or shrink in a similar manner to crystal clusters in the gas phase.
Recalling the results described in \S \ref{origin}, we assume a system of 1D bosons made up of a mixture of different boson $n$-mers as shown in Fig. \ref{fermions}.
We refer to this inhomogeneous picture of a vicinal surface as the step-droplet picture.

Let us now consider such a many-body system of boson $n$-mers (Fig. \ref{fermions}).
We denote the density of giant steps with height $n$ by $\rho_n$.
We assume that the interactions among $n$-mers are short range.
The free energy $f_n(\rho_n)$ of a vicinal surface made up of such giant steps then becomes the GMPT universal type since
\beq
f_n(\rho_n)= f(0)+\gamma_n (\phi) \rho_n + B_n (\phi)  \rho_n^3 + C_n (\phi)  \rho_n^4 + {\cal{O} }(\rho_n^5), \label{fn}
\eeq
where $\gamma_n (\phi) $ represents the step tension of a giant step, $B_n (\phi) $ represents the step interaction coefficient among giant steps, $C_n (\phi) $ ($<0$) represents the three-body interaction coefficient among giant steps\cite{cn}, and $\phi$, similar to the case for a single step (Fig. \ref{diagonal_line}), represents the angle between the mean running direction of the giant step and the $y$-axis.
In the limit $\rho_n \rightarrow 0$, $\phi$ converges to $\phi_0$.

Recalling that $\rho_n = N_{s,n}/L=|\vec{p}|/d_n =|\vec{p}|/(n d_1)$, where $N_{s,n}$ is the total number of giant steps with size $n$, $L$ is the length of the projected area of the vicinal surface, and $d_n$ is the height of the giant steps, we obtain a $|\vec{p}|$-expanded expression for $f_n(\rho_n)$ as $f_n(\vec{p})$ (\ref{derivation}, Eq. (\ref{fp-n})).
We then expand $\gamma_n(\phi)$ and $B_n(\phi)$ with respect to $n$ around $n=1$.
After some calculations  (\ref{derivation}), we obtain 
\beqa
%f_n(\rho_n)=f_n(|\vec{p}|/d_n) \nonumber\\
f_n(|\vec{p}|)= f(0)+\left[\gamma_1 (\phi)  -\gamma_1^{(1)} (\phi) +\frac{\gamma_1^{(2)} (\phi) }{2} +
 (\gamma_1^{(1)} (\phi) -\gamma_1^{(2)} (\phi) )n \right. \nonumber \\
\left. 
+ \frac{\gamma_1^{(2)} (\phi) }{2}n^2 \right] \frac{|\vec{p}|}{d_1} %\nonumber \\
+ \frac{B_1 (\phi) }{n^4} \frac{|\vec{p}|^3}{d_1^3} \left[1+ (n-1)\frac{\tilde{\gamma}_1^{(1)} (\phi) }{\tilde{\gamma}_1 (\phi) }
+\frac{1}{2}(n-1)^2 \frac{\tilde{\gamma}_1^{(2)} (\phi) }{\tilde{\gamma}_1 (\phi) }     
\right]^{-1}\nonumber \\
+ \frac{C_n (\phi) }{n^4}\frac{ |\vec{p}|^4}{d_1^4} + {\cal{O}} (|\vec{p}|^5), \label{fpn2}
\eeqa
where  $\gamma_1^{(m)} (\phi)$ and $\tilde{\gamma}_1^{(m)} (\phi)$ are defined by
\beq
\gamma_1^{(m)} (\phi) =\left. \frac{\partial^m (\gamma_n (\phi) /n)}{\partial n^m}\right|_{n=1}, \quad \tilde{\gamma}_1^{(m)} (\phi) =\left. \frac{\partial^m (\tilde{\gamma}_n (\phi) /n)}{\partial n^m}\right|_{n=1},
\eeq
and $\tilde{\gamma}_n (\phi)  = \gamma_n (\phi)  + \partial^2 \gamma_n (\phi)  / \partial \phi^2$ represents the stiffness of a giant step.

Since $n$ is difficult to estimate, we approximate Eq. (\ref{fpn2}) by replacing $n $ with $\langle n (\phi) \rangle$.
Recalling the results for $\langle n (\phi) \rangle$ obtained by the Monte Carlo calculations in \S \ref{mean}, we expand $\langle n (\phi) \rangle$ with respect to $|\vec{p}|$ around $|\vec{p}|=0$.
Namely,
\beqa
\langle n(\phi) \rangle = 1+ n_0^{(1)} (\phi) |\vec{p}|+ \frac{1}{2}n_0^{(2)} (\phi) |\vec{p}|^2+ \frac{1}{6}n_0^{(3)} (\phi) |\vec{p}|^3 + \cdots \nonumber \\
\left. n_0^{(m)} (\phi) = \frac{\partial^m \langle n (\phi) \rangle }{\partial |\vec{p}|^m}\right|_{|\vec{p}|=0+} . \label{n_expand}
\eeqa

Finally, substituting Eq. (\ref{n_expand}) into Eq. (\ref{fpn2}), and after some calculations,  the non-GMPT vicinal surface free energy can finally be expressed as 
\beqa
f_{\rm eff}(\vec{p}) \equiv f_{\langle n \rangle}(\vec{p})=f(0)+\gamma_1 (\phi)  \frac{|\vec{p}|}{d_1}+A_{\rm eff} (\phi)  |\vec{p}|^2 \nonumber \\
+B_{\rm eff} (\phi)  |\vec{p}|^3 +C_{\rm eff} (\phi)  |\vec{p}|^4+{\cal O} (p^5), \label{fpeff}
\eeqa
where
\beqa
A_{\rm eff} (\phi) =n_0^{(1)} (\phi) \gamma_1^{(1)} (\phi) /d_1 \label{eqAeff} \\
B_{\rm eff} (\phi) =  \frac{1}{2d_1} \left[ n_0^{(2)} (\phi) \gamma_1^{(1)} (\phi) + n_0^{(1)} (\phi)^2 \gamma_1^{(2)} (\phi) \right] + \frac{B_1 (\phi) }{d_1^3}   \label{eqBeff} \\
C_{\rm eff} (\phi) = \frac{1}{6d_1}\left[ n_0^{(3)} (\phi) \gamma_1^{(1)} (\phi)  +  3 n_0^{(1)} (\phi) n_0^{(2)} (\phi) \gamma_1^{(2)}  (\phi)  \right] \nonumber \\
- \frac{B_1 (\phi)  n_0^{(1)} (\phi) }{d_1^3}  \left( 4  +  \frac{\tilde{\gamma}_1^{(1)} (\phi) }{\tilde{\gamma}_1 (\phi) }\   \right) 
+ \frac{C_n (\phi) }{d_1^4}  . \label{eqCeff}
\eeqa
In Eq. (\ref{fpeff}), the term $|\vec{p}|^2$ appears.
In addition, $B_{\rm eff}(\phi)$ can be larger or smaller than $B_1(\phi)$. 

The microscopic quantities $\epsilon$ and $\epsilon_{\rm int}$ do not appear explicitly in Eq. (\ref{fpeff}).
They do, however, seem to determine mesoscopic quantities such as step tension, stiffness, and mean height.
The mean step height depends on the character of the step-step attraction (see Fig. \ref{nfig}(a)).
In fact, the slope dependence of the mean step height for the RSOS-I model seems to be different to that for the p-RSOS model.
A detailed study on the RSOS-I model will be reported elsewhere\cite{akutsu11-2}.

\subsection{Thermodynamic expression for non-GMPT shape exponents}

In order to derive the non-GMPT shape exponents thermodynamically, we write the $|\vec{p}|$-expanded form of the vicinal surface free energy as
\beq
f_{g}(\vec{p})= f(0) +\gamma(\phi) |\vec{p}|+ B_{\zeta}(\phi)|\vec{p}|^{\zeta} + O(|\vec{p}|^{\zeta+1}),
\label{extend_f}
\eeq
where $\zeta>1$.
We can then express the shape exponents thermodynamically as (\ref{a_exponents})
\beqa
\theta_t= 2 \zeta/(\zeta-1), \quad  {\cal A}_t(\phi)= \frac{1}{\theta_t} \left[\frac{(\kBT)^{\zeta +1}}{2 \zeta B_{\zeta}(\phi)\tilde{\gamma}(\phi)^{\zeta}} \right]^{ \frac{1}{\zeta-1}} , 
 \label{exponent_y}
\eeqa
for the shape exponent along the tangential line ($\overline{QQ'}$ in Fig. \ref{diagonal_line}), and 
\beqa
\theta_n = \zeta/(\zeta-1), \quad  {\cal A}_n(\phi)= \frac{1}{\theta_n} \left[\frac{\kBT}{\zeta B_{\zeta}(\phi)} \right]^{ \frac{1}{\zeta-1}} . 
\label{exponent_x}
\eeqa
for the shape exponent along the normal line ($\overline{PP'}$ in Fig. \ref{diagonal_line}).
From Eq. (\ref{exponent_y}) and Eq. (\ref{exponent_x}), we obtain the following relationships:
\beqa
\theta_t=2 \theta_n \label{scaling}\\
r = \left[ \frac{{\cal A}_t(\phi)}{{\cal A}_n(\phi)} \right]^{\frac{1}{\theta_n}} t^2 = \frac{\kBT}{2 \tilde{\gamma}(\phi)}t^2. \label{radius}
\eeqa

\section{Application to p-RSOS model}

\subsection{$T_{f,2}< T \leq T_{f,1}$}

\begin{figure}%[!htb]
\begin{center}
\includegraphics[width=9 cm,clip]{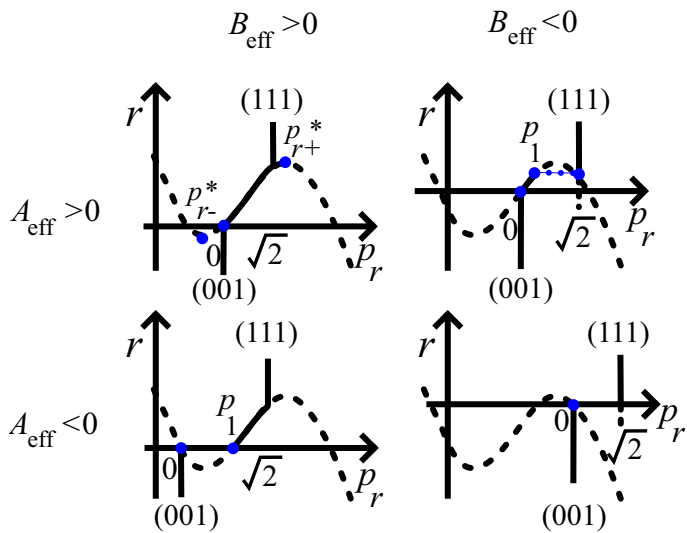}
\end{center}
\caption{\label{rp}
Schematic diagram of slope dependence of $r$.
Solid lines:  $r(p_r)$ of the ECS.
Broken lines: Formally obtained $r(p_r)$ using Eq. (\ref{fpeff}).
$p^*_{r \pm}$: Slopes which give extrema of $r(p_r)$ (Eq. (\ref{p*eq})). 
}
\end{figure}

We compare the results of Eqs. (\ref{exponent_y}) and (\ref{exponent_x}) for the non-GMPT free energy Eq. (\ref{fpeff}) with the numerical results in \S 2 obtained by statistical mechanical calculations.

Fig. \ref{rp} is a schematic diagram of the $r$-$p_r$ curve, where $C_{\rm eff}(\phi)$ is assumed to be negative, $r$ is given by $r= \partial f_{\rm eff}(\vec{p})/\partial p_r$, and $p_r$ is the surface slope along the line $\overline{PP'}$ (Fig. \ref{diagonal_line}).
Let us define $p_{r\pm}^*$ so that $\partial r(p_r)/\partial p_r =0$.
We then have
\beq
p_{r\pm}^*=\frac{1}{12 C_{\rm eff}(\phi)}\left[ -3 B_{\rm eff}(\phi)
\pm \sqrt{9B_{\rm eff}(\phi)^2 - 24 A_{\rm eff}(\phi) C_{\rm eff}(\phi)}\right]. \label{p*eq}
\eeq
%where we assume the angle dependence of $A_{\rm eff}$, $B_{\rm eff}$, and $C_{\rm eff}$ implicitly.
%We abbreviate the notation with respect to the angle $\phi$, for a while. %
When $A_{\rm eff}(\phi)>0$ and $B_{\rm eff}(\phi)<0$, for $0 \leq p <p_1$, a vicinal surface with a regular train of steps is in an equilibrium state, for $p_1<p<p_{r+}^* $, it is metastable, and for $p_{r+}^*<p<p_0$ ($p_0=\sqrt{2}$), it is thermodynamically unstable.

For the temperature region $T_{f,2}< T \leq T_{f,1}$, we have $n_0^{(1)} (\pi/4)>0$ in \S \ref{mean} and $\gamma_1^{(1)}(\pi/4) \geq 0$ in \S \ref{non-univ-exponents}, so that $A_{\rm eff}(\pi/4)\geq 0$.
From Eq. (\ref{exponent_y}) and Eq. (\ref{exponent_x}), we have
\beq
\theta_n=2, \quad \theta_t= 4,  \quad  {\cal A}_n= \frac{\kBT}{4A_{\rm eff}(\pi/4)}, \quad  {\cal A}_t= \frac{(\kBT)^3}{16A_{\rm eff}(\pi/4)\tilde{\gamma}_1(\pi/4)^2}. \label{nonuniv_exp}
\eeq

By comparing Eq. (\ref{nonuniv_exp}) with the PWFRG results in \S \ref{non-univ-exponents}, we see that the non-universal shape exponents for the p-RSOS model agree with the values obtained using Eq. (\ref{nonuniv_exp}).
Moreover, by comparing the amplitudes of Eq. (\ref{nonuniv_exp}) with the PWFRG results in \S \ref{non-univ-exponents}, we obtain $\beta A_{\rm eff}(\pi/4)= (6.22 \pm 0.06) \times 10^{-3}$ and $\beta \tilde{\gamma}_1(\pi/4)=1.39\pm 0.03$.
This value of $\beta \tilde{\gamma}_1(\pi/4)$ agrees well with  $\beta \tilde{\gamma}(\pi/4)_{\rm Ising }$ (Eq. (\ref{stiff11})).
Consequently, in the $p_r \rightarrow 0$ limit, step droplets ``evaporate'' and dissociate into individual steps with $n=1$. 
Substituting the value of $\beta A_{\rm eff}(\pi/4)$ and the value of $n_0^{(1)}= 2.26$ obtained from Fig. \ref{nfig} into Eq. (\ref{eqAeff}), we have $\beta \gamma_1^{(1)}(\pi/4) = (2.75 \pm 0.05) \times 10^{-3}$.
The small value of $\beta \gamma_1^{(1)}(\pi/4)$  is consistent with experimental observations on Si(113)\cite{sudoh}.

\subsection{$T>T_{f,1}$}

From the results in \S \ref{non-univ-exponents}, a vicinal surface with small $p_r$ shows the GMPT universal behavior of single steps.
Near $p_r \approx 0$, therefore, $A_{\rm eff}(\pi/4)=0$, and then $\gamma_1^{(1)}(\pi/4) $ is considered to be zero.
Hence, we have from Eq. (\ref{exponent_x}), Eq. (\ref{exponent_y}) and Eq. (\ref{fpeff}), 
\beqa
\theta_n=3/2, \quad \theta_t=2\theta_n=3,  \label{univ_exp1}\\
p_x=p_r/\sqrt{2}=\sqrt{\frac{\kBT}{6 B_{\rm eff}(\pi/4)}} r^{1/2}, \nonumber \\ %\quad  
|p_x-p_y|=\sqrt{2} p_t=\frac{(\kBT)^2}{\sqrt{3B_{\rm eff}(\pi/4)\tilde{\gamma}_1(\pi/4)^3}} |t|^2.\label{univ_pr1}  
\eeqa
By comparing Eq. (\ref{univ_pr1}) with the curve calculated using the PWFRG method, we have $\beta B_{\rm eff}(\pi/4) = 0.281 \pm 0.008$.
$\beta B_1(\pi/4)$ at $\kBT/\epsilon=0.37$ is estimated as  $\beta B_1(\pi/4) = 1.206 $ by use of the step stiffness of the Ising model (Eq. (\ref{stiff11})).
Then, from Eq. (\ref{eqBeff}) with $n_0^{(1)}=1.45$ (Fig. \ref{nfig}) and with $\gamma_1^{(1)}(\pi/4) = 0$, we have $\beta \gamma_1^{(2)}(\pi/4)= -0.88 \pm 0.08$.

\subsection{$T \leq T_{f,2}$}

In the low temperature region, the first-order shape transition occurs at the (001) facet edge on the ECS profile.
If we consider $A_{\rm eff}(\pi/4)<0$ and $B_{\rm eff}(\pi/4)<0$ (Fig. \ref{rp}), the first-order shape transition can be understood in terms of the non-GMPT vicinal surface free energy $f_{\rm eff}(\vec{p})$ (Eq. (\ref{fpeff})).
From the EFS results (\S \ref{efs}), we have $\gamma_1^{(1)}(\pi /4)<0$, and from the Monte Carlo calculation of $\langle n \rangle$ at $\kBT/\epsilon = 0.35$, we have $n_0^{(1)}>0$.
These results are consistent with $A_{\rm eff}(\pi/4)<0$.

\section{Summary and discussion}

In \S2, we present statistical mechanical calculations on the vicinal surface of the restricted solid-on-solid (RSOS) model with a point-contact type step-step attraction (p-RSOS model) (Fig. \ref{vicinal}).
Applying the product wave-function renormalization group (PWFRG) method, which is a variant of the density matrix renormalization group (DMRG) method, to the transfer matrix for the p-RSOS model, we calculate the reduced equilibrium crystal shape (ECS) of the p-RSOS model (Fig. \ref{ECS3D}) and the surface gradient $\vec{p}=(p_x,p_y)$ as a function of $X$ ($= \beta \eta_x= - \beta \lambda x$, $\beta = 1/\kBT$) and $Y$ ($= \beta \eta_y= - \beta \lambda y$) (Fig. \ref{peta110}).

We obtain the first-order shape transition around the (111) facet on the ECS profile for $T<T_{f,1}$.
For $T<T_{f,2}$, a first-order shape transition at the (001) facet edge is also observed on the ECS profile, where the (001) facet directly contacts the (111) facet (Fig. \ref{EFS}).
 By analyzing the PWFRG results, we obtain the equilibrium facet shape (Fig. \ref{EFS}) and the step tension; for the vicinal surface tilted towards the $\langle 110 \rangle$ direction, we obtain the step stiffness, the non-GMPT shape exponents (Fig. \ref{exponent}) and the non-GMPT amplitudes (Fig. \ref{p_eta_detail}).

In \S 3, in order to elucidate the origin of the non-GMPT behavior, we study step droplets formed by thermal step bunching on the vicinal surface tilted towards the $\langle 110 \rangle$ direction for the p-RSOS model.
We calculate the vicinal surface free energy from the Andreev surface free energy obtained from the PWFRG calculations (Fig. \ref{bfpp}).
 Step bunching near equilibrium caused by a singularity in the surface free energy (Fig. \ref{fig_hill_and_valley08}) is demonstrated using Monte Carlo simulations with a simple Metropolis algorithm (Fig. \ref{mcsfigLowT}, \ref{mcsfig}).
To obtain a clear image of local step droplets, we demonstrate giant step formation similar to step faceting for $T \leq T_{f,2}$.
From a microscopic point of view, these giant steps are formed from sticky steps due to the step-step attraction.
For $T_{f,2} \leq T \lesssim T_{f,1}$, the giant steps partially dissociate due to entropic repulsion.
Using a long-duration Monte Carlo simulation, we calculate the slope dependence of the mean step number $\langle n \rangle$ in a giant step (Fig. \ref{nfig}), and find that $\langle n \rangle$ increases linearly with $|\vec{p}|$.

In \S 4, we derive a $|\vec{p}|$-expanded form of the non-GMPT vicinal surface free energy $f_{\rm eff}(\vec{p})$ (Eq. (\ref{fpeff})).
In Eq. (\ref{fpeff}), the $|\vec{p}|^2$ term appears.
In the derivation, the concept of step droplets (boson $n$-mers, Fig. \ref{fermions}) and knowledge of the slope dependence of $\langle n \rangle$ obtained using the long-duration Monte Carlo calculation in \S \ref{mean} are crucial. 

In \S 5, the results in \S 2 are consistently reproduced by thermodynamical calculations based on the non-GMPT vicinal surface free energy $f_{\rm eff}(\vec{p})$.
By comparing the results in \S 2 with the results obtained using $f_{\rm eff}(\vec{p})$, we obtain information on the derivative of $\gamma_n$. $\partial (\gamma_n/n) /\partial n|_{n=1} $ is small and positive in the temperature range $T_{f,2} \leq T \lesssim T_{f,1}$, whereas it is negative for $T<T_{f,2}$ and zero for $T>T_{f,1}$. $\partial ^2 (\gamma_n/n) /\partial n^2|_{n=1} <0$ for $T \approx T_{f,1}$ and the step droplets are dispersed in the limit $|\vec{p}| \rightarrow 0$.

The first-order transition on the ECS profile has been studied theoretically by several authors\cite{rottman84,jayaprakash84-2}.
Rottman and Wortis\cite{rottman84} calculated the interface tension using a 3D cubic Ising model with both nn and next nearest neighbor (nnn) interactions between spins by means of the mean field approximation.
They found the first-order shape transition around the (001) facet at low temperature for the negative nnn interactions. 
However, they did not discuss the shape exponent for the temperature range $T_3<T<T_t$ (using their terminology).
Jayaprakash {\it et al.}\cite{jayaprakash84-2} studied the vicinal surface of the interacting terrace-step-kink (TSK) model with long-range step-step attractions corresponding to attractive dipolar interactions.
 They determined the surface free energy using mean field calculations, and showed that the surface free energy $f(\vec{p})$ has the form of the GMPT (Eq. (\ref{frho})), and that the long-range attractions changes $B$.
When $B$ becomes negative, the facet edge causes the first-order shape transition.
This explanation for the first-order shape transition is limited to the case of attractive long-range step-step interactions, because in the case of short-range attractions, $B$ does not change.
Hence, the explanation based on step droplets described in \S \ref{non-gmpt} is required.

In real systems, there are a variety of different surface effects that occur, as described in \S \ref{faceting}.
Of these, elastic interactions among steps are the most important source of long-range step-step repulsion. 
The TSK model with a long-range step-step repulsion of the order of $\sim 1/l^2$, where $l$ is the inter-step distance, is known to have the GMPT universal free energy\cite{jayaprakash84-2,yamamoto94} in a homogeneous system, and the system has the GMPT universal shape exponents on the ECS.
The work of Shenoy {\it et al.}\cite{shenoy98} suggests that step bunching occurs as $m$-mers of particles in the TSK model with a long-range step-step repulsion of the order of $\sim 1/l^2$ and a short-range step-step attraction.
In this case, if $m$ depends on $\vec{p}$, then the non-GMPT shape exponents on the ECS are expected from Eq. (\ref{fpeff})-Eq. (\ref{eqCeff}).

Paulin {\it et al.}\cite{paulin,misba10} considered a surface system with an elastic interaction of $\sim 1/l^2 \ln l$ in order to study the dynamical behavior of the single-step to double-step transition on the Si(100) surface. 
Such an elastic interaction is thought to change the absolute values of the step tension, stiffness, and interaction coefficient (using our terminology), as in the case of $\sim 1/l^2$ \cite{yamamoto94}.
As long as a homogeneous system is realized (``in phase meandering'' using their terminology), an elastic interaction of $\sim 1/l^2 \ln l$ probably does not change the shape exponents.
However, in the case of ``out of phase meandering'', since inhomogeneity is suggested, non-GMPT shape exponents are expected due to the step droplets as explained in \S \ref{non-gmpt}.

Recently, a first-order like shape change for $^4$He around the (0001) facet has been reported\cite{parshin11}. 
The strong anisotropy in the surface stiffness measured in the experiment suggests the existence of a singularity in the surface free energy.
The results of the present study and those in our previous reports\cite{akutsu09,akutsu10} are consistent with this finding.

\section{Conclusion}
In the present study, we investigate a vicinal surface tilted towards the $\langle 110 \rangle$ direction near a (001) facet on the ECS of the p-RSOS model.
For $T_{f,2}<T \lesssim T_{f,1}$, the shape exponents on the ECS have non-GMPT values such as $\theta_n =2$ for the normal direction and $\theta_t=4$ for the tangential direction.
The origin of the non-GMPT shape exponents is the formation of step droplets (``giant steps'') with different sizes.
The non-GMPT expression for the vicinal surface free energy $f_{\rm eff}(\vec{p})$ is derived.
In the derivation, knowledge of the $|\vec{p}|$ dependence of the mean step-droplet size is crucial.
The results obtained by statistical mechanical calculations using the p-RSOS model are successfully reproduced using thermodynamical calculations based on $f_{\rm eff}(\vec{p})$.

\section{Acknowledgements}

The author would like to thank Prof. T. Yamamoto and Prof. Y. Akutsu for discussions.
This work was supported in part by  the ``Research for the Future'' Program of The Japan Society for the Promotion of Science (JSPS-RFTF97P00201) and by a Grant-in-Aid for Scientific Research from the Ministry of Education, Science, Sports and Culture (No. 15540323).

\appendix

\section{Exact expressions for interface tension and interface stiffness on a two-dimensional square Ising model\label{2dising}}

The exact expression for the 2D ECS for a nn square Ising model\cite{akutsu90}-\cite{akutsu99iop} is written using the angle $\phi$ in Fig. \ref{diagonal_line}(b) as follows:
\beqa
D(X_c,Y_c)=0, \nonumber \\
 \frac{ \partial D(X,Y)}{\partial Y}|_{(X,Y)=(X_c,Y_c)}= \tan \phi \ \frac{\partial D(X,Y)}{\partial X}|_{(X,Y)=(X_c,Y_c)}, \label{Dfunction}
\eeqa
where $D(X,Y)$ is
\beq
D(X,Y)=\cosh (X)+\cosh (Y) - \frac{\cosh^2 ( \beta \epsilon)}{\sinh ( \beta \epsilon)}. \label{Dfunction1}
\eeq
Hence, the equation for the ECS on the 2D square nn Ising model becomes 
\beq
\cosh (X_c)+\cosh (Y_c) = \frac{\cosh^2 ( \beta \epsilon)}{\sinh ( \beta \epsilon)}. \label{ecs_Ising1}
\eeq

Let us denote the anisotropic interface (step) tension for an interface with a mean running direction parallel to the line $\overline{QQ'}$ by $\gamma (\phi)$ (Fig. \ref{diagonal_line}), and let us consider the interface stiffness  $\tilde{\gamma}(\phi)=\gamma(\phi)+\partial^2\gamma(\phi)/\partial \phi^2$.
%\subsection{Ising model calculation of the step tension and the step stiffness}
The exact expression for the interface stiffness for the 2D nn Ising model is derived by use of $D(X,Y)$  (Eq. (\ref{Dfunction})) with Eq. (\ref{eqgam}), ${\rm d} D(X(\phi),Y(\phi))/{\rm d} \phi|_{(X,Y)=(X_c,Y_c)}=0$ and ${\rm d}^2 D(X(\phi),Y(\phi))/{\rm d} \phi^2|_{(X,Y)=(X_c,Y_c)}=0$ as follows:
\beqa
\beta \tilde{\gamma}(\phi)_{\rm Ising}= \left[ \frac{\partial D(X,Y)}{\partial X} \cos \phi 
+ \frac{\partial D(X,Y)}{\partial Y} \sin \phi \right] \nonumber \\
\times \left[ \frac{\partial^2 D(X,Y)}{\partial X^2} \sin^2 \phi 
+ \frac{\partial^2 D(X,Y)}{\partial Y^2} \cos^2 \phi \right. \nonumber \\
\left. \left. -2 \frac{\partial^2 D(X,Y)}{\partial Y \partial X} \sin \phi \cos \phi  \right]^{-1} \right|_{(X,Y)=(X_c,Y_c)}. \label{stiff_phi}
\eeqa
\section{Derivation of the non-GMPT vicinal surface free energy\label{derivation}}

We investigate the $n$ dependence of $f_n(\rho_n)$ for $\rho_n \approx 0$.
$\rho_n$ is expressed as $\rho_n = N_{s,n}/L=|\vec{p}|/d_n $, where $N_{s,n}$ is the total number of giant steps with size $n$,  $L$ is the length of the projected area of the vicinal surface, and $d_n$ is the height of the giant step.
$L$ is assumed to be sufficiently large so that we may take the thermodynamic limit.
Substituting the expressions for $\rho_n$ and $d_n=n d_1$ into Eq. (\ref{fn}), we have
\beq
f_n(\vec{p})= f(0)+\frac{\gamma_n (\phi) }{d_1 n} |\vec{p}| + \frac{B_n (\phi) }{d_1^3 n^3} |\vec{p}|^3 + \frac{C_n (\phi) }{d_1^4 n^4} |\vec{p}|^4 + {\cal{O}} (p^5). \label{fp-n}
\eeq
Since $\gamma_n (\phi) $ is considered to be approximately equal to $n\gamma_1 (\phi) $, we expand $\gamma_n (\phi) /n$ with respect to $n$ around $n=1$ as follows:
\beqa
\frac{\gamma_n (\phi) }{n}= \gamma_1 (\phi)  +\gamma_1^{(1)} (\phi)  (n-1)  +\frac{1}{2}\gamma_1^{(2)} (\phi)  (n-1)^2+ \cdots \nonumber \\
\left. \gamma_1^{(m)} (\phi) =\frac{\partial^m (\gamma_n (\phi) /n)}{\partial n^m}\right|_{n=1} \label{gamma_n'}
\eeqa
Assuming the universal relation Eq. (\ref{univrel}), where $\tilde{\gamma}_n(\phi)$ represents the stiffness of the giant step $\tilde{\gamma}_n (\phi)  = \gamma_n (\phi)  + \partial^2 \gamma_n (\phi)  / \partial \phi^2$, we expand $\tilde{\gamma}_n (\phi)$ instead of $B_n (\phi)$.
That is,
\beqa
\frac{\tilde{\gamma}_n (\phi) }{n}= \tilde{\gamma}_1 (\phi)  +\tilde{\gamma}_1^{(1)} (\phi)  (n-1)  +\frac{1}{2}\tilde{\gamma}_1^{(2)} (\phi)  (n-1)^2+ \cdots \nonumber \\
\left. \tilde{\gamma}_1^{(m)} (\phi) =\frac{\partial^m (\tilde{\gamma}_n (\phi) /n)}{\partial n^m}\right|_{n=1}.\label{tilde_gamma_n'}
\eeqa%
Substituting Eq. (\ref{gamma_n'}) and Eq. (\ref{tilde_gamma_n'}) into Eq. (\ref{fp-n}), we have Eq. (\ref{fpn2}).

\section{Thermodynamic derivation of the shape exponents\label{a_exponents}}

\begin{figure}%[htbp]
\begin{center}
%\epsfysize= 3 cm
%\centerline{\epsfbox{fig7akutsu061v1eps7.eps}}
\includegraphics[width=10cm,clip]{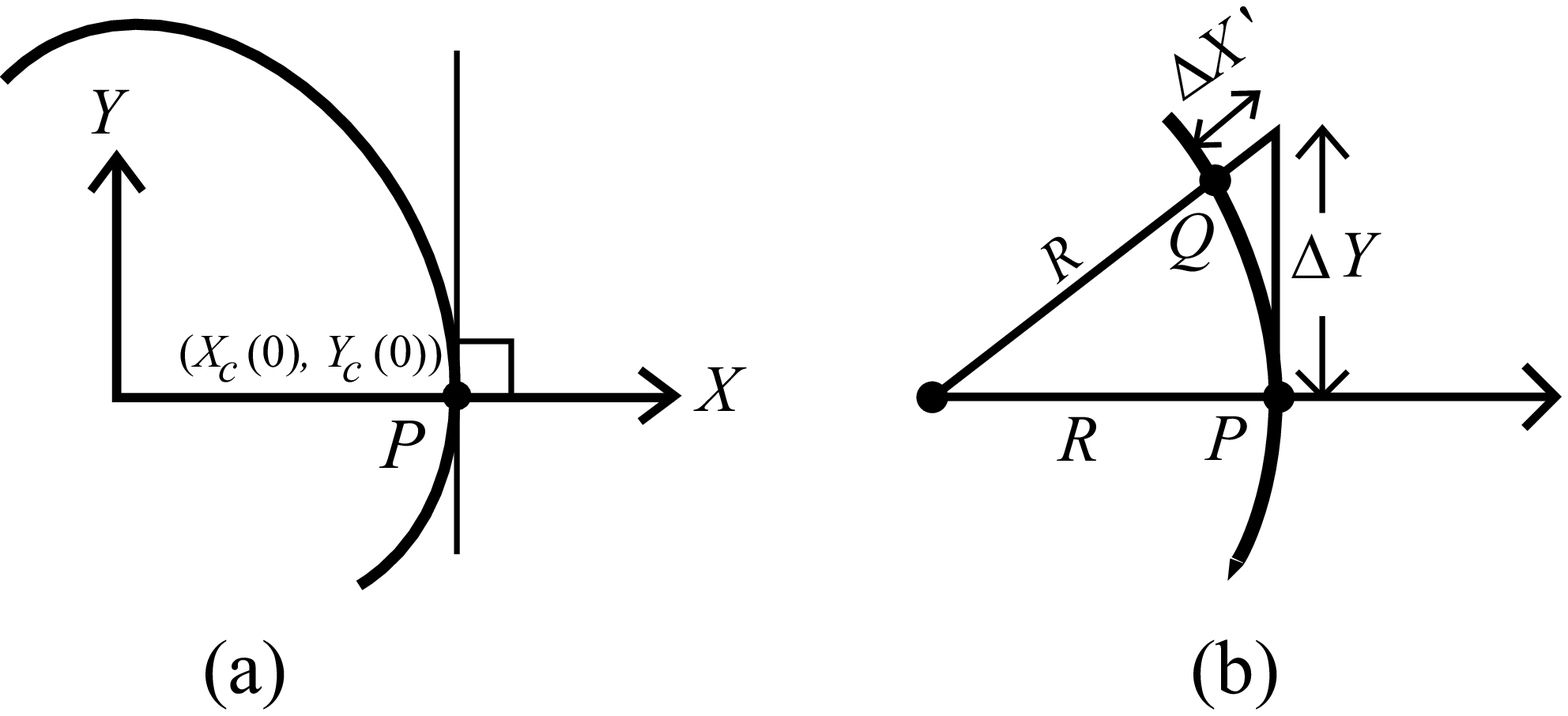}%
\caption{
(a) Choice of $X$- and $Y$-axes at $P$ on the facet contour.
(b) $Q$ lies near $P$ on the facet contour.
 The ``normal distance'' $\Delta X'$ is related to ``tangential distance'' $\Delta
 Y$ as $\Delta X'=(\Delta Y)^2/2R$, where $R$ is the radius of curvature at $P$.
}
\label{choice}
\end{center}
\end{figure}

Eq. (\ref{exponent_y}) - Eq. (\ref{radius}) are derived from the extended vicinal surface free energy $f_g(\vec{p})$ (Eq. (\ref{extend_f})).
In the following manner, we consider the case where the first-order shape transition does not occur on the contour of the (001) facet.

From the thermodynamics of the ECS\cite{andreev}, the coordinates $(X,Y)$ on the reduced ECS are obtained from $f(\vec{p})$ as follows\cite{akutsu87-2}:
\beq
X=\beta \frac{\partial f(\vec{p})}{\partial p_x},
 \quad Y=\beta \frac{\partial f(\vec{p})}{\partial p_y}, \label{XYdef}
\eeq
where $\beta =1/\kBT$.
Then the reduced coordinates $(X, Y)$ on the ECS are written 
\beqa
X&=&X_c(\phi)+|\vec{p}|^{\zeta-1} \left[ \beta \zeta B_{\zeta}(\phi) \cos\phi - \beta B_{\zeta}'(\phi)\sin\phi \right],
 \nonumber \\
Y&=&Y_c(\phi )+|\vec{p}|^{\zeta-1}\left[ \beta \zeta B_{\zeta}(\phi ) \sin\phi+ \beta B_{\zeta}'(\phi ) \cos\phi \right] , \nonumber \\
&&B_{\zeta}'(\phi )=\partial B_{\zeta}(\phi )/\partial \phi , \label{tag8}
\eeqa
where $X_c$ and $Y_c$ are given as follows:
\beqa
X_c&=& \beta \gamma(\phi) \cos \phi -\beta \gamma' (\phi) \sin \phi, \nonumber \\
Y_c&=& \beta \gamma(\phi) \sin \phi +\beta \gamma' (\phi) \cos \phi ,\nonumber \\
&&\gamma' (\phi)= \partial \gamma (\phi)/\partial \phi. \label{eqefs_gmpt} 
\eeqa

Let us choose the $X$- and $Y$- axes so that the $Y$-axis is parallel to the tangential line of the facet contour at $P$ (Fig. \ref{choice}).
 With this choice of coordinate system, we have $\phi=0$
 at $P$ and $\gamma'(0)=0$ (Eq. (\ref{eqefs_gmpt})).  
 Along the tangential line ($X=X_c(0)=0$), $\phi$ and $|\vec{p}|$ are
 not independent but are constrained to satisfy
\beq
-\frac{1}{2}\tilde{\gamma}(0)\phi^2+\zeta B_{\zeta}(0)|\vec{p}|^{\zeta-1} =0,
\label{tag12}
\eeq
which is derived by expanding (\ref{tag8}) and (\ref{eqefs_gmpt}) with
 respect to $\phi$ and $|\vec{p}|$ ($|\vec{p}|<<1$ and $|\phi|<<1$, near the point $P$).  Combining (\ref{tag12}) with (\ref{tag8}) and (\ref{eqefs_gmpt}),
 we obtain
\beq
\phi =\frac{\kBT \Delta Y}{\tilde{\gamma}(0)},\ 
|\vec{p}|=\left[ \frac{(\kBT)^2 \Delta Y^2}{2 \zeta \tilde{\gamma}(0)B_{\zeta}(0)}\right]^{\frac{1}{\zeta-1}}   \label{tag13}
\eeq
along the tangential line $\overline{PQ}$ in Fig. \ref{choice}, where $\Delta Y = Y-Y_c$.
 Near $P$, $Z=\tilde{f}(X_c(0),\Delta Y+Y_c(0))/\kBT$
 is expanded to give
\begin{eqnarray}
\Delta Z &=& - \frac{ [\gamma(0)+ \frac{1}{2}\gamma''(0)\phi^2]}{\kBT}|\vec{p}| + \frac{B_{\zeta}(0)}{\kBT}|\vec{p}|^{\zeta}
 \nonumber \\
   & &- \Delta Y |\vec{p}|\phi -X_c(0)|\vec{p}|(1-\frac{1}{2}\phi^2),
 \label{tag14}
\end{eqnarray}
where $\Delta Z = Z(X,Y)-Z(X_c,Y_c)$.
Substituting (\ref{tag13}) into (\ref{tag14}), we obtain expressions for the tangential shape exponent $\theta_t$ and the tangential amplitude ${\cal A}_t(\phi)$ as follows:
\beqa
\Delta Z=-{\cal A}_t(0)|\Delta Y|^{\theta_t}, \nonumber \\
{\cal A}_t(0)= \frac{1}{\theta_t} \left[\frac{(\kBT)^{\zeta +1}}{2 \zeta B_{\zeta}(0)\tilde{\gamma}(0)^{\zeta}} \right]^{ \frac{1}{\zeta-1}} , \quad
\theta_t= 2 \zeta/(\zeta-1). %\ref{exponent_y}
\eeqa
Similarly, along the normal line ($Y-Y_c(0)=0$), we obtain expressions for the normal amplitude ${\cal A}_n(\phi)$ and the normal shape exponent $\theta_n$ as follows:
\beqa
\Delta Z=-{\cal A}_n(0)(\Delta X)^{\theta_n}, \quad (X_c(0)\leq X) \nonumber \\
{\cal A}_n(0)= \frac{1}{\theta_n} \left[\frac{\kBT}{\zeta B_{\zeta}(0)} \right]^{ \frac{1}{\zeta-1}} , \quad
\theta_n = \zeta/(\zeta-1). %\ref{exponent_x}
\eeqa

From Eq. (\ref{exponent_y}) and Eq. (\ref{exponent_x}), we obtain a scaling relation like
\beq
\theta_t=2 \theta_n %\ref{scaling}
\eeq
when $\tilde{\gamma}(0) \neq 0$.
From Eq. (\ref{exponent_x}) and Eq. (\ref{exponent_y}), we obtain the relation
\beq
\Delta X = \left[ \frac{{\cal A}_t(0)}{{\cal A}_n(0)} \right]^{\frac{1}{\theta_n}}\Delta Y^2 = \frac{\kBT}{2 \tilde{\gamma}(0)}\Delta Y^2. %\ref{radius}
\eeq
This expression is also obtained geometrically from Fig. \ref{choice}(b) using the radius of curvature $R$ such that $\Delta X' = \Delta Y^2/ (2 R) = \kBT \Delta Y^2/ (2 \tilde{\gamma}(0))$\cite{akutsu86} and $\Delta X'= \Delta X$ for small $\Delta X'$ and $\Delta Y$.


\begin{thebibliography}{999}
\bibitem{cn} Note that $C_n (\phi) $ ($<0$) is caused by the RSOS restriction.
The existence of the higher order terms in the original RSOS system is confirmed numerically by Fig. \ref{peta110}(b).
\end{thebibliography}
\end{document}